\documentclass[acmsmall]{acmart}

\AtBeginDocument{%
  }

\setcopyright{acmlicensed}
\copyrightyear{2026}
\acmYear{2026}
\acmDOI{XXXXXXX.XXXXXXX}

\acmJournal{CSUR}
\acmVolume{1}
\acmNumber{1}
\acmArticle{1}
\acmMonth{8}

\usepackage{amsfonts}
\usepackage{algorithmic}
\usepackage{algorithm}
\usepackage{subcaption}
\usepackage{stfloats}
\usepackage{url}
\usepackage{verbatim}
\usepackage{siunitx}
\usepackage[utf8]{inputenc}
\usepackage{enumitem}
\usepackage{tikz}
\usepackage[acronym]{glossaries}

\usetikzlibrary{shapes.geometric, arrows.meta, positioning}

\glsdisablehyper % disable glossary links

\begin{document}

% The cover letter is usually submitted separately in ACM journals, 
% but kept here as per your original structure.
%\input{content/cover_letter}
\newpage

%\title{A Comprehensive Survey of Redundancy Systems: Techniques, Applications, and Future Trends with a Focus on Triple Modular Redundancy (TMR)}
\title{A Comprehensive Survey of Redundancy Systems with a Focus on Triple Modular Redundancy (TMR)}

\author{Lukas Flad}
\email{lukas.flad@th-koeln.de}
\affiliation{%
  \institution{University of Applied Sciences Cologne}
  \department{Faculty of Information, Media and Electrical Engineering}
  \city{Cologne}
  \country{Germany}
}

\author{Mark Leyer}
\email{mark.leyer@th-koeln.de}
\affiliation{%
  \institution{University of Applied Sciences Cologne}
  \department{Faculty of Information, Media and Electrical Engineering}
  \city{Cologne}
  \country{Germany}
}

\author{Felix Sebastian Nitz}
\email{felix.nitz@th-koeln.de}
\affiliation{%
  \institution{University of Applied Sciences Cologne}
  \department{Faculty of Information, Media and Electrical Engineering}
  \city{Cologne}
  \country{Germany}
}

\author{Tobias Krawutschke}
\email{tobias.krawutschke@th-koeln.de}
\affiliation{%
  \institution{University of Applied Sciences Cologne}
  \department{Faculty of Information, Media and Electrical Engineering}
  \city{Cologne}
  \country{Germany}
}

%% By default, the full list of authors will be used in the page headers.
\renewcommand{\shortauthors}{Flad et al.}

%% article. It must be placed BEFORE \maketitle.
\begin{abstract}
Despite its maturity, the field of fault-tolerant redundancy suffers from significant terminological fragmentation, where functionally equivalent methods are frequently described under disparate names across academic and industrial domains. This survey addresses this ambiguity by providing a structured and comprehensive analysis of redundancy techniques, with a primary focus on Triple Modular Redundancy (TMR). A unified taxonomy is established to classify redundancy strategies into Spatial, Temporal, and Mixed categories, alongside the introduction of a novel five-class framework for voter architectures. Key findings synthesize practical trade-offs, contrasting high-reliability spatial TMR for safety-critical applications against resource-efficient temporal methods for constrained systems. Furthermore, the shift toward Mixed and Adaptive TMR (e.g., Approximate Triple Modular Redundancy (ATMR), X-Rel) for dynamic and error-tolerant applications, such as Artificial Intelligence (AI) acceleration, is explored. This work identifies critical research gaps, including the threat of Multi-Bit Upsets (MBUs) in sub-28nm technologies, the scarcity of public-domain data on proprietary high-integrity systems, and the absence of high-level toolchains for dynamic reconfiguration. Finally, suggestions are offered for future research directions, emphasizing the need for terminological standardization, MBU-resilient design methodologies, and the development of open-source tools for adaptive fault tolerance.
\end{abstract}

\begin{CCSXML}
<ccs2012>
   <concept>
       <concept_id>10010583.10010750.10010751.10010757</concept_id>
       <concept_desc>Hardware~System-level fault tolerance</concept_desc>
       <concept_significance>500</concept_significance>
       </concept>
 </ccs2012>
\end{CCSXML}

\ccsdesc[500]{Hardware~System-level fault tolerance}

\keywords{Triple Modular Redundancy (TMR), N-Modular Redundancy (NMR), Fault Tolerance, Hardware TMR, Software TMR, Adaptive TMR, AI-Assisted TMR, Voter Logic, Multi-Bit Upsets, Radiation-Hardened Systems, Aerospace Applications, FPGA Redundancy, Redundant Multi-Threading, Reliability Metrics, System-Level Triplication, Single Event Upset (SEU)}

\maketitle

\newacronym{TMR}{TMR}{Triple Modular Redundancy}
\newacronym{MTMR}{MTMR}{Modified Triple Modular Redundancy}
\newacronym{AI}{AI}{Artificial Intelligence}
\newacronym{SEU}{SEU}{Single-Event Upset}
\newacronym{FPGA}{FPGA}{Field-Programmable Gate Array}
\newacronym{NMR}{NMR}{N-Modular Redundancy}
\newacronym{DMR}{DMR}{Dual/Double Modular Redundancy}
\newacronym{SoC}{SoC}{System on a Chip}
\newacronym{MCU}{MCU}{Microcontroller}
\newacronym{MBU}{MBU}{Multiple-bit Upset}
\newacronym{LUT}{LUT}{Lookup Table}
\newacronym{SRAM}{SRAM}{Static random-access memory}
\newacronym{CNN}{CNN}{Convolutional Neural Network}
\newacronym{I/O}{I/O}{Input/Output}
\newacronym{MAC}{MAC}{Multiply-Accumulate}
\newacronym{NMOS}{NMOS}{n-type metal-oxide-semiconductor}
\newacronym{LET}{LET}{Linear energy transfers}
\newacronym{CMOS}{CMOS}{complementary metal-oxide-semiconductor}
\newacronym{ASIC}{ASIC}{application-specific integrated circuit}
\newacronym{RoRA}{RoRA}{Reliability-oriented Place \& Route Algorithm}
\newacronym{ALU}{ALU}{Arithmetic Logic Unit}
\newacronym{IP}{IP}{Intellectual Property}
\newacronym{QMR}{QMR}{Quadruple Modular Redundancy}
\newacronym{EASA}{EASA}{European Union Aviation Safety Agency}
\newacronym{ANN}{ANN}{Artificial Neural Networks}
\newacronym{RMT}{RMT}{Redundant Multi-Threading}
\newacronym{TMT}{TMT}{Temporal Multi-Threading}
\newacronym{MTTF}{MTTF}{mean time to failure}
\newacronym{RISC-V}{RISC-V}{Reduced Instruction Set Computer\,V}
\newacronym{IMT}{IMT}{interleaved multi-threading}
\newacronym{WCET}{WCET}{worst-case execution time}
\newacronym{MTBF}{MTBF}{mean time between failures}
\newacronym{CRC}{CRC}{ Cyclic Redundancy Check}
\newacronym{CERN}{CERN}{Conseil Européen pour la Recherche Nucléaire (European Organization for Nuclear Research)}
\newacronym{FDI}{FDI}{Fault Detection and Isolation}
\newacronym{SDC}{SDC}{Silent-Data Corruption}
\newacronym{WNR}{WNR}{Weapons Neutron Research Facility}
\newacronym{MWTF}{MWTF}{Mission-Weighted Time to Failure}
\newacronym{ATMR}{ATMR}{Approximate Triple Modular Redundancy}
\newacronym{FAC}{FAC}{Fault-tolerant Approximate Computing}
\newacronym{H-TMR}{H-TMR}{Hierarchical Triple Modular Redundancy}
\newacronym{RHBD}{RHBD}{Radiation-Hardened by Design}
\newacronym{RCN}{RCN}{Reconfiguration Control Networks}
\newacronym{TMRSB}{TMRSB}{Triple Modular Redundancy with Standby}
\newacronym{ICAP}{ICAP}{Internal Conﬁguration Access Port}
\newacronym{API}{API}{Application Programming Interface}
\newacronym{RTL}{RTL}{Register Transfer Level}
\newacronym{FTMR}{FTMR}{Functional Triple Modular Redundancy}
\newacronym{AFTR}{AFTR}{Adaptive Fault-Tolerant Re-Execution}
\newacronym{PB}{PB}{Primary-Backup}
\newacronym{PE}{PE}{Primary-Exception}
\newacronym{SCP}{SCP}{Self-Configuring Pair}
\newacronym{AFT}{AFT}{Adaptive Fault Tolerance}
\newacronym{X-Rel}{X-Rel}{Approximate Reliability}
\newacronym{MTED}{MTED}{Maximum Tolerable Error Distance}
\newacronym{MSSIM}{MSSIM}{Mean Structural Similarity Index}
\newacronym{EDAP}{EDAP}{Energy-Delay-Area Product}
\newacronym{IDMR}{IDMR}{Inexact Dual Modular Redundancy}
\newacronym{ITDMR}{ITDMR}{Inexact Triple Modular Redundancy}
\newacronym{ILP}{ILP}{Integer Linear Programming}
\newacronym{LSB}{LSB}{Least Significant Bit}
\newacronym{MSB}{MSB}{Most Significant Bit}
\newacronym{PSNR}{PSNR}{Peak Signal-to-Noise Ratio}
\newacronym{SSIM}{SSIM}{Structural Similarity Index Measure}
\newacronym{FD}{FD}{Fault Detection}
\newacronym{DHR}{DHR}{Dual Hardware Redundancy}
\newacronym{TPMC}{TPMC}{Three-Phase Matrix Converter}
\newacronym{IGBT}{IGBT}{Insulated Gate Bipolar Transistor}
\newacronym{ECC}{ECC}{error-correcting codes}
\newacronym{RMS}{RMS}{Root Mean Square}
\newacronym{STROBES}{STROBES}{Software-Level Triple Modular Redundancy for On-Board Embedded Systems}
\newacronym{COTS}{COTS}{Commercial Off-The-Shelf}
\newacronym{RT}{RT}{Real-Time}
\newacronym{Zynq}{Zynq}{Zynq System-on-Chip}
\newacronym{FFNN}{FFNN}{Feedforward Neural Network}
\newacronym{TSPC}{TSPC}{True Single-Phase Clock}
\newacronym{SET}{SET}{Single Event Transient}
\newacronym{DSP}{DSP}{Digital Signal Processing}
\newacronym{DWC}{DWC}{Dual Modular Redundancy with Comparison}
\newacronym{CUDA}{CUDA}{Compute Unified Device Architecture}
\newacronym{GPU}{GPU}{Graphics Processing Unit}
\newacronym{ASIL}{ASIL}{Automotive Safety Integrity Level}
\newacronym{ROTS}{ROTS}{Real-time operating system}
\newacronym{TL-NMR}{TL-NMR}{task-level N modular redundancy}
\newacronym{SEL}{SEL}{Single Event Latchup}
\newacronym{SCR}{SCR}{silicon-controlled rectifier}
\newacronym{ECU}{ECU}{Electronic Control Unit}
\newacronym{SOI}{SOI}{Silicon on Insulator}
\newacronym{McellU}{MCU}{Multi-cell Upset}
\newacronym{SBU}{SBU}{Single-Bit Upset}
\newacronym{ORGA}{ORGA}{Optically Reconfigurable Gate Array}
\newacronym{SEFI}{SEFI}{Single Event Functional Interrupts}
\newacronym{TID}{TID}{Total Ionization Dose}
\newacronym{QDI}{QDI}{quasi delay insensitive}
\newacronym{CPU}{CPU}{central processing unit}
\newacronym{OS}{OS}{Operating System}
\newacronym{DTMR}{DTMR}{dynamic TMR}
\newacronym{FFT}{FFT}{Fast Fourier Transform}
\newacronym{TTMR}{TTMR}{Time-Triple Modular Redundancy}
\newacronym{PFC}{PFC}{Primary Flight Computer}
\newacronym{MCC}{MCC}{Multi-Change Controller}
\newacronym{ISO}{ISO}{International Organization for Standardization}
\newacronym{CDC}{CDC}{Clock Domain Crossing}
\newacronym{EDC}{EDC}{Error Detection and Correction}
\newacronym{LTMR}{LTMR}{Localized TMR}
\newacronym{STMR}{STMR}{Selective TMR}
\newacronym{RHIC}{RHIC}{Relativistic Heavy Ion Collider}

\section{Introduction}
The demand for dependable computing has become a critical design pillar across all major technology sectors. As systems in aerospace, automotive, and data-critical applications grow in complexity, their vulnerability to transient and permanent faults increases. Ensuring operational reliability—the ability to function correctly despite such faults—is paramount. This survey provides a comprehensive analysis of redundancy, the foundational design strategy to achieve this fault tolerance.

\subsection{Motivation}
The field of fault-tolerant information processing system design offers a wide array of redundancy techniques and voter architectures. This challenge is not merely academic. In safety-critical domains, selecting a suboptimal or misunderstood redundancy technique can lead to wasted resources, certification failures, or catastrophic system failure. While this diversity is a sign of maturity, it also creates significant complexity. Naming inconsistencies, overlapping concepts, and fragmented terminology—where identical methods are often described by different names—make comparative assessment and selection particularly challenging, directly hindering the transfer of knowledge from research to practice.

\textbf{Goal:} This survey directly confronts these issues by providing a structured, analytical, and comprehensive overview of redundancy and voter techniques. It consolidates existing work, clarifies terminology, and—most critically—\textbf{introduces a novel, method-oriented categorization} for both redundancy schemes and voter architectures. By creating a unified taxonomy, this paper provides an indispensable guide for researchers and practicing engineers, demystifying the complex landscape and enabling the design of more robust, reliable, and efficient systems.

\begin{figure}[htbp]
    \centering
    \includegraphics[width=\textwidth]{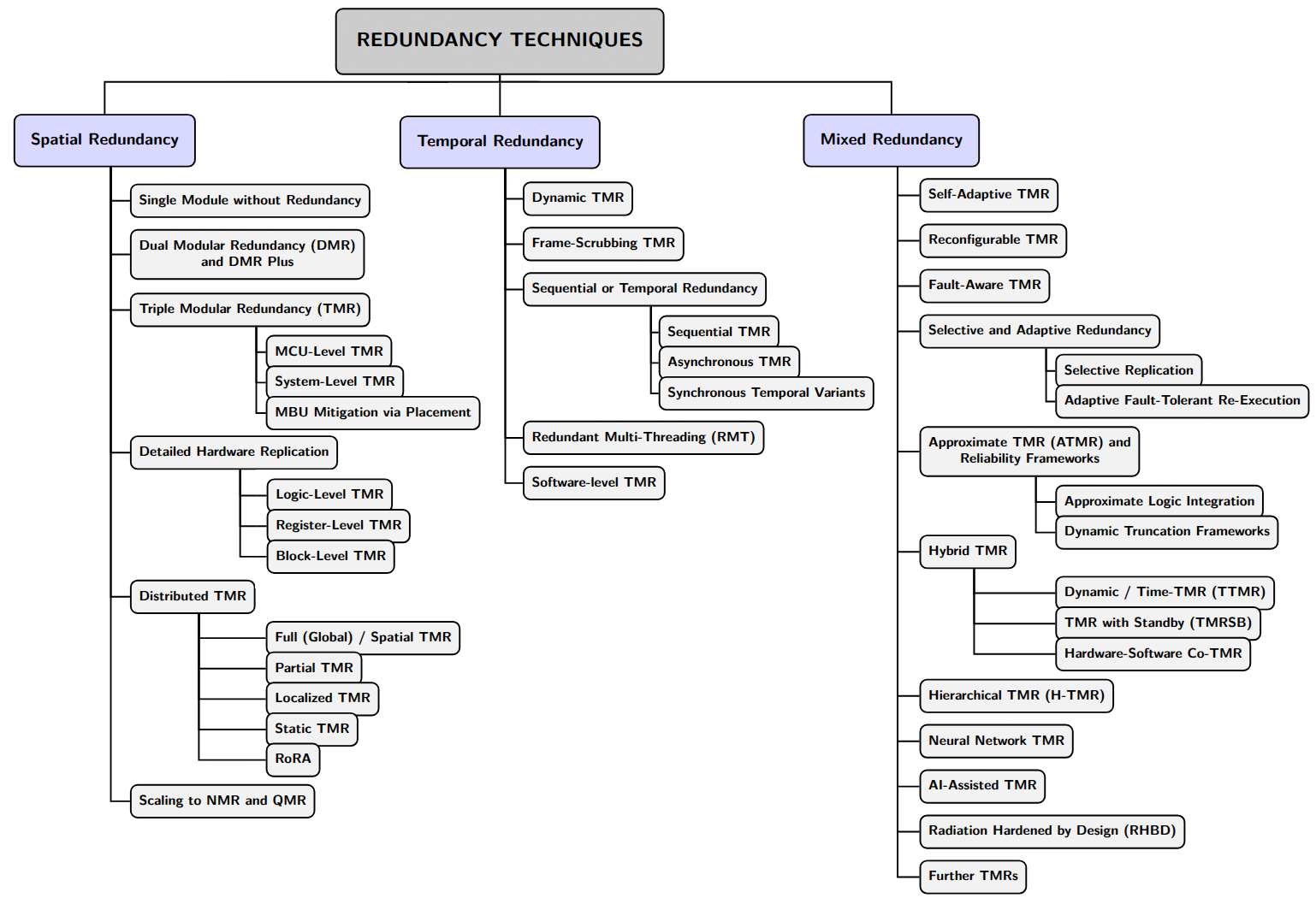}
    \Description{A hierarchical tree diagram classifying redundancy techniques into three main branches: Spatial Redundancy, Temporal Redundancy, and Mixed Redundancy. Each branch expands into numerous subcategories, detailing specific methodologies like DMR, TMR, Distributed TMR, and Hybrid techniques.}
    \caption{Comprehensive taxonomy of redundancy techniques. This high-level overview illustrates the scale of the design space, which is analytically broken down in subsequent sections.}
    \label{fig:redundancy_tree}
\end{figure}

\begin{figure}[htbp]
    \centering
    \includegraphics[width=\textwidth]{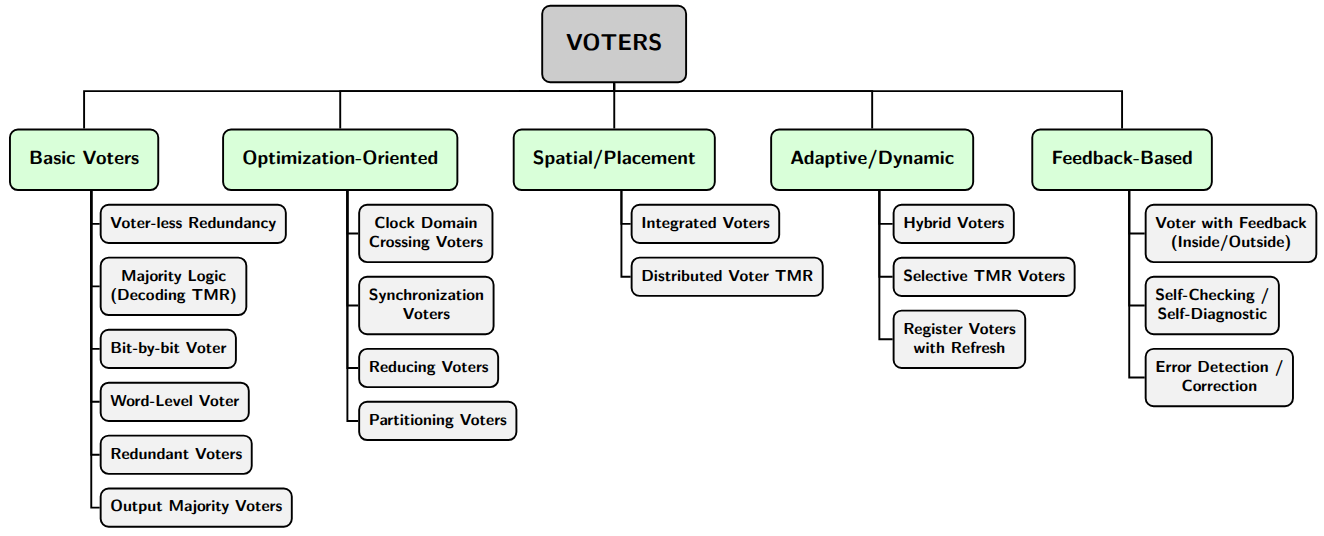}
    \Description{A hierarchical tree diagram classifying voter architectures into five main branches: Basic Voters, Optimization-Oriented, Spatial/Placement, Adaptive/Dynamic, and Feedback-Based. Each branch details specific voter implementations, such as Word-Level, Synchronization, Distributed, and Self-Checking voters.}
    \caption{Novel, method-oriented categorization of voter architectures. The structured branches provide a unified framework to navigate the diverse array of voter implementations.}
    \label{fig:voter_tree}
\end{figure}

To visually illustrate the sheer scale and diversity of the fault-tolerant design space, Figure \ref{fig:redundancy_tree} and Figure \ref{fig:voter_tree} map out the comprehensive taxonomies of redundancy strategies and voter architectures developed in this survey. While this vast array of methodologies can initially appear overwhelming—reflecting the complexity and fragmentation previously noted—this survey is specifically structured to systematically break down these extensive trees into digestible, logically organized categories in the subsequent chapters.

\subsection{Structure}

This work is divided into four major chapters:

\begin{itemize}
    \item \textbf{Fundamentals} — Introduces core principles and terminology surrounding redundancy and voters.
    \item \textbf{Analytical Methods} — Presents a classification of redundancy and voter mechanisms, organized to guide the reader structurally and logically through the design space.
    \item \textbf{Discussion and Key Learnings} --- Offers comparative insights, limitations, and application-specific recommendations, along with a conclusion and future work.
\end{itemize}

The analytical classification of electronic redundancy techniques is organized into three conceptual layers—spatial, temporal, and mixed—based on recurring taxonomies in the literature~\cite{pakstas_redundancy_2001, ding_classification_2017}. The \emph{mixed} category is introduced to account for emerging or hybrid methods that would otherwise cause overlap between classification branches, improving readability and usability.

In contrast to redundancy, the classification of voter architectures is much less developed in existing research. Although individual publications discuss specific implementations—such as the Word-Voter by Mitra and McCluskey~\cite{mitra_word-voter_2000} and the self-checking voter design by Afzaal and Lee~\cite{afzaal_self-checking_2018}—these works address only narrow aspects of voter logic and do not provide a unified taxonomy. To the best of the authors’ knowledge, no comprehensive classification of voter designs exists in the current literature. As such, the structural categorization presented here—comprising basic, optimization-oriented, spatial/placement, adaptive/dynamic, and feedback-based voter types—has been newly developed in the context of this survey. It is intended to offer a systematic and practical framework for comparing and selecting voter architectures in fault-tolerant systems.

To identify relevant literature, a structured search was conducted using three major academic databases: IEEE, SpringerLink, and Google Scholar. The search included combinations of the keywords \textit{redundancy, fault tolerance, TMR, NMR, voter logic, spatial redundancy, temporal redundancy, and fault-masking}. To ensure accessibility for future research, only publicly and readily available sources were included at the time of writing.

The literature selection strategy was twofold: to balance foundational, high-impact research with the current state-of-the-art. Foundational works were identified by their high citation counts and established relevance. Concurrently, recent publications were included to capture emerging and novel techniques, even if not yet widely cited. This flexible approach ensures the inclusion of both seminal papers and specific, novel implementations, such as those in:
\begin{enumerate}[label=\textbf{\arabic*.}, wide, labelindent=0pt]
    \item AI-assisted voting~\cite{ahmad_modified_2025},
    \item dynamic fault tolerance~\cite{barbirotta_evaluation_2022}, and
    \item approximate reliability frameworks~\cite{vafaei_x-rel_2023}.
\end{enumerate}
This inclusion strategy is intended to support reproducibility and accessibility while enabling readers to efficiently explore the landscape of redundancy and voter techniques.

By building on this rigorous and flexible literature foundation, this paper offers a unique and comprehensive perspective. This is the first survey to systematically categorize both redundancy techniques and voter architectures into a unified, practical framework. This work aims to serve as a new foundational reference for the field.

From here on out, whenever it is referred to redundancy, information processing redundancy is implicit. Some techniques could also apply to mechanical redundancy, but the focus of this survey is solely on electronics, software and information processing systems.

\section{Related Work}
While redundancy and fault-tolerant design have been extensively studied \cite{reddy_fault-tolerant_2000, rahimi_fault_2020, avizienis_basic_2004, avizienis_n-version_1985, shannon_probabilistic_1956, pratt_improving_2006, mitra_word-voter_2000, ganesan_redundancy_2017, lyons_use_1962,arifeen_approximate_2020} over the past decades, there is currently no comprehensive survey that systematically categorizes both redundancy techniques and voter architectures in an integrated and structured form. Existing literature often focuses on specific aspects—such as hardware-level \gls{TMR}~\cite{habinc_sandi_functional_2002}, software-based N-version programming~\cite{avizienis_n-version_1985}, or fault-tolerant voter design~\cite{mitra_word-voter_2000}—but lacks a unified view across multiple abstraction layers and implementation domains.

This survey addresses that gap by consolidating established and emerging approaches—including spatial, temporal, and hybrid redundancy schemes—alongside a taxonomy of voter architectures. It aims to assist practitioners and researchers in identifying suitable fault-tolerance strategies without the need to consult dozens of isolated sources.

\section{Fundamentals} 

Redundancy is a foundational concept in dependable system design, aimed at enhancing reliability and fault tolerance. This chapter introduces the core principles of redundancy with a particular emphasis on \gls{TMR} (see Section \ref{subsec:Fundamentals_TMR}. As systems grow in complexity—especially in safety-critical fields like aerospace and automotive—an understanding of redundancy techniques becomes essential.

The idea of redundancy, in the context of computers, dates back decades~\cite{avizienis_n-version_1985} and has been applied across various domains to improve system reliability and prevent single-point failures. While redundancy can appear in many technical forms, its essence lies in replicating system-critical components and functions to ensure continuity during faults. \gls{TMR} is the most widely used redundancy model especially in radiation-prone environments, like space systems~\cite{habinc_sandi_suitability_2002, shao_research_2024, sekioka_radiation-hardened_2025, sterpone_analysis_2005, quinn_domain_2007}. Consequently, this paper focuses primarily on \gls{TMR} as the most representative redundancy scheme in both theoretical and applied contexts.

Rather than merely duplicating components, modern redundancy strategies are embedded into the architecture, enabling fault masking, self-diagnosis, and graceful degradation~\cite{habinc_sandi_functional_2002, anjankar_fault_2020}. These methods aim to prevent complete system breakdowns while ensuring continued data correctness and output validity.

% Add purpose statement for paper
This survey serves as a foundational reference for understanding, comparing, and selecting redundancy mechanisms—eliminating the need for practitioners to analyze hundreds of technical papers. It provides structured guidance to match redundancy techniques with specific system requirements and constraints.

\subsection{What is Redundancy?}

Redundancy refers to the intentional \emph{N}-times replication of critical system elements—whether components, logic paths, or functional modules—to maintain correct system operation in the presence of faults.

Its goals can be broadly divided into two categories:
\begin{itemize}
    \item \textbf{Functional robustness}: enables continued operation even when some modules fail~\cite{ anjankar_fault_2020, rahimi_fault_2020}.
    \item \textbf{Data safety}: ensures integrity of internal and output data despite transient or permanent errors~\cite{afzaal_self-checking_2018, habinc_sandi_functional_2002}.
\end{itemize}

Redundancy is vital for \textbf{fail-operational behavior}, where systems continue functioning under fault conditions, rather than entering a fail-silent or fail-stop state. This is critical for sectors such as \gls{FPGA} in radiation prone environments such as aerospace, particle accelerators such as \gls{CERN}, the \gls{RHIC} or more basic in automotive applications~\cite{milluzzi_exploration_2017, bertoa_fault-tolerant_2023}.

It enables fault tolerance techniques like majority voting, error detection, and isolation strategies~\cite{ding_classification_2017}. These not only improve reliability but also simplify safety certification by making faults detectable and predictable.

\subsection{Triple Modular Redundancy} \label{subsec:Fundamentals_TMR}

\gls{TMR} is a prominent fault tolerance method that replicates a system module three times. All three modules perform the same task in parallel, and a majority \textbf{voter} selects the correct output. If one of the three components fails, the voter can still determine the correct result from the remaining two~\cite{afzaal_self-checking_2018, habinc_sandi_functional_2002}.

\gls{TMR} is emphasized throughout this survey because it represents a widely adopted, well-balanced fault-tolerance strategy across critical domains such as aerospace, automotive, and radiation-hardened computing \cite{avizienis_n-version_1985, habinc_sandi_functional_2002, yeh_triple-triple_1996, stoffel_-demand_2024}. Unlike \gls{DMR}, which allows for fault detection but lacks intrinsic error correction, \gls{TMR} offers in-line fault masking via majority voting, thereby enabling uninterrupted operation in the presence of single-point failures \cite{reviriego_dmr_2016}. While higher-order \gls{NMR}, such as 5MR or \gls{QMR}, can theoretically handle more faults, the associated increase in selective protection logic, area overhead, and voter complexity often yields diminishing practical returns \cite{el-maleh_generalized_2014}.
In short: higher costs for little to no gains over \gls{TMR}. Therefore, among all redundancy strategies, \gls{TMR} is predominantly favored due to its optimal trade-off between system reliability, implementation complexity, and resource overhead.

As the primary focus in this study is \gls{TMR}, many of the redundancy concepts, voter designs, and architectural principles discussed herein may also extend, and thereby are not limited to \gls{DMR}, \gls{NMR}, and hybrid schemes in suitable contexts.

\subsection{Voters in TMR Systems}

Voters play a central role in redundancy techniques such as \gls{TMR} architectures. They aggregate outputs and determine the most likely correct value using majority logic. Depending on system constraints, voters may be implemented in hardware or software, and they can vary in complexity—ranging from simple bit-level to full word-based designs~\cite{rahimi_fault_2020, afzaal_self-checking_2018, mitra_word-voter_2000}. Voter design affects the overall reliability; for example, single-voter failures can become system-wide single points of failure unless mitigated via self-checking logic or redundant voting circuits~\cite{reviriego_dmr_2016}.

\subsection{Recent Developments in TMR}

Recent advancements in \gls{TMR} focus on minimizing area and energy overheads while improving fault coverage. Variants like \gls{MTMR}, partial TMR, or adaptive and AI-assisted TMR have emerged to optimize reliability–performance trade-offs~\cite{sanchez-clemente_partial_2016, gonzalez_adaptive_1997, bertoa_fault-tolerant_2023}.

In addition, \gls{FPGA} and spaceborne systems have adopted hybrid redundancy models that combine functional replication with self-checking or reconfiguration features~\cite{habinc_sandi_suitability_2002, afzaal_self-checking_2018}. These developments mark a shift from static to dynamic and intelligent fault-tolerance schemes.

\section{Methods} \label{sec:methodes} 

This chapter provides a structured classification of redundancy strategies and voter architectures in fault-tolerant systems.
The methods are divided into two central categories:

\begin{itemize}
    \item \textbf{Redundancy}: techniques to ensure functional and data safety through replication or timing-based tolerance, organized into spatial, temporal, and mixed redundancy.
\item \textbf{Voters}: decision mechanisms that select or infer correct outputs from redundant instances, categorized into five major classes.
\end{itemize}

Each technique is described in terms of its structural concept, relevance, and known variants.
Several methods are functionally identical but named differently across publications; these are included to support comprehensive literature referencing.
This paper, however, preferentially uses the most common and frequently cited terminology.
This classification was developed specifically for this survey in order to address the lack of a consistent and complete redundancy taxonomy in existing literature.
While earlier works—such as those by Avizienis~\cite{avizienis_n-version_1985} and Pakštas~\cite{pakstas_redundancy_2001}—introduced foundational structures, they do not reflect the full range of techniques found in modern systems, nor do they adequately separate conceptual overlaps.
In particular, the \textbf{mixed redundancy} category has been introduced in this work to resolve ambiguous cases that do not clearly fit solely into either spatial nor temporal classifications.
This enables a more readable and logically consistent methodology chapter and supports a step-by-step (tree-like exploration) orientation for readers seeking to match redundancy techniques to system-level requirements, application constraints, or failure models.
\subsection{Redundancy} \label{sec:redundancy}

As discussed before (see section~\ref{sec:methodes}) redundancy strategies will be classified into the following three types:

\begin{itemize}
    \item \textbf{Spatial Redundancy}: achieves fault tolerance by distributing redundant components across physical space.
This includes full system replication (e.g. \gls{TMR}), but also encompasses placement-aware design to prevent co-location of redundant logic—critical in mitigating shared-fault vulnerabilities such as radiation-induced upsets~\cite{kastensmidt_optimal_2005, yeh_triple-triple_1996, pratt_improving_2006}.
\item \textbf{Temporal Redundancy}: operates on the principle of time-based re-execution.
Fault tolerance is achieved by repeating computations across time slots rather than replicating hardware resources.
Examples include time-multiplexed \gls{TMR} or re-scheduled execution threads~\cite{chen_task_2016, barbirotta_evaluation_2022, she_time_2009}.
\item \textbf{Mixed Redundancy}: combines spatial and temporal methods or includes schemes that cannot be clearly assigned to either.
This includes self-adaptive, reconfigurable, or AI-enhanced techniques which shift dynamically between modes based on system state or environment~\cite{afzaal_self-checking_2018, chu_fault_1990, ding_classification_2017}.
\end{itemize}

\begin{figure}[htbp]
    \centering
    \includegraphics[width=0.55\linewidth]{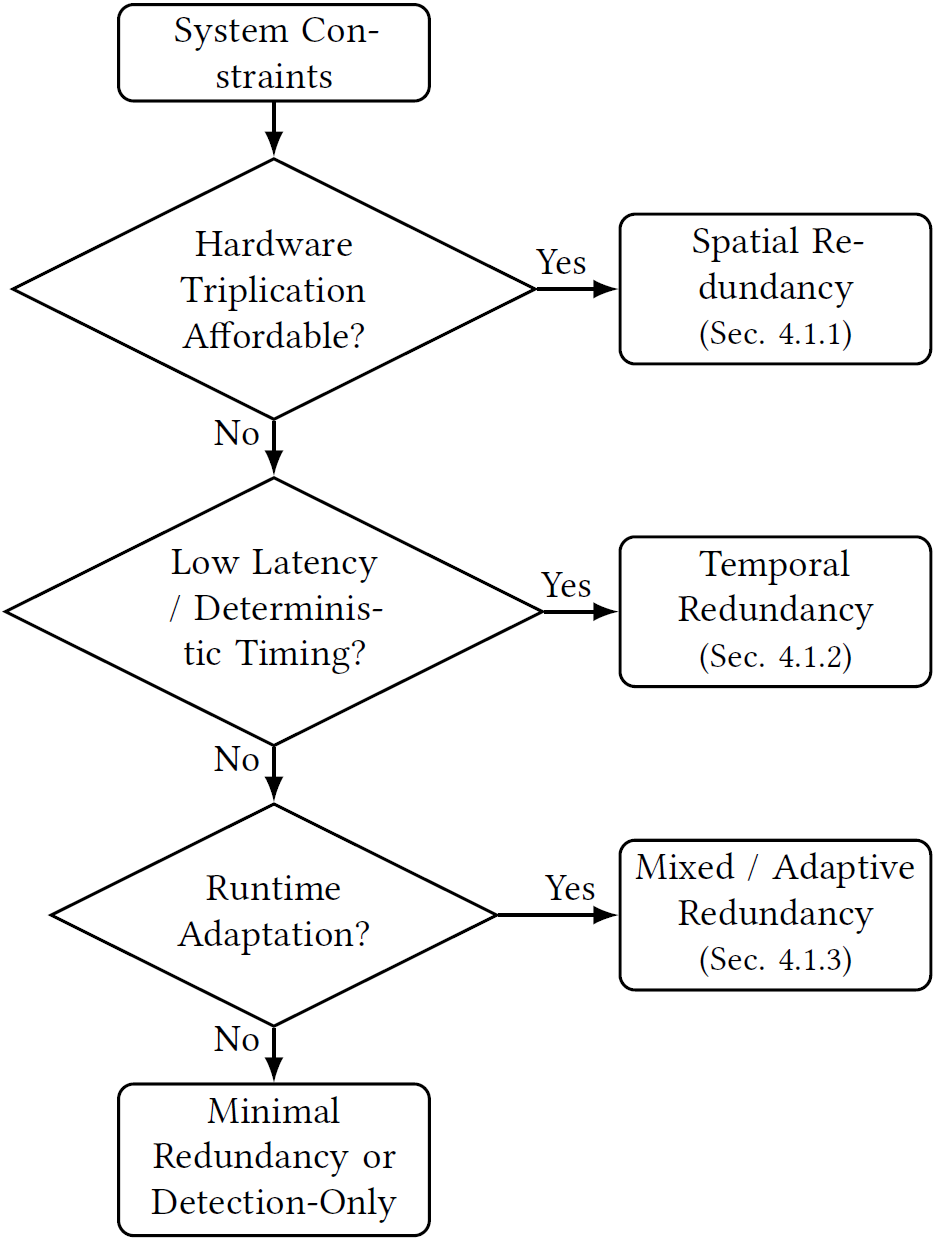}
    \caption{User-centered decision tree for navigating redundancy strategies.}
    \Description{A flowchart starting with a block labeled 'System Constraints'. An arrow points down to a decision diamond 'Hardware Triplication Affordable?'. If Yes, an arrow points right to a block 'Spatial Redundancy'. If No, an arrow points down to a decision 'Low Latency / Deterministic Timing?'. If Yes, it points right to 'Temporal Redundancy'. If No, it points down to a decision 'Runtime Adaptation?'. If Yes, it points right to 'Mixed / Adaptive Redundancy'. If No, it points down to a final block 'Minimal Redundancy or Detection-Only'.}
    \label{fig:redundancy_navigation_tree}
\end{figure}

\paragraph{Decision Guidance (Fig.~\ref{fig:redundancy_navigation_tree})}
As an overview, Fig.~\ref{fig:redundancy_navigation_tree} presents the redundancy navigation tree, guiding the reader through the taxonomy of spatial, temporal, and mixed redundancy techniques.
This helpful guideline can be used to be directed into the desired topic and functionality of redundancy.
Each subsection outlines key design principles, representative methods, and relevant variations.
Where applicable, equivalent techniques are grouped to avoid duplication, with naming diversity documented for clarity.
%--------------------------------------------------------------------
% SPATIAL REDUNDANCY
%--------------------------------------------------------------------

\subsubsection{Spatial Redundancy} \label{sec:spatial_redundncy}

\begin{figure}[htbp]
    \centering
    \includegraphics[width=0.55\linewidth]{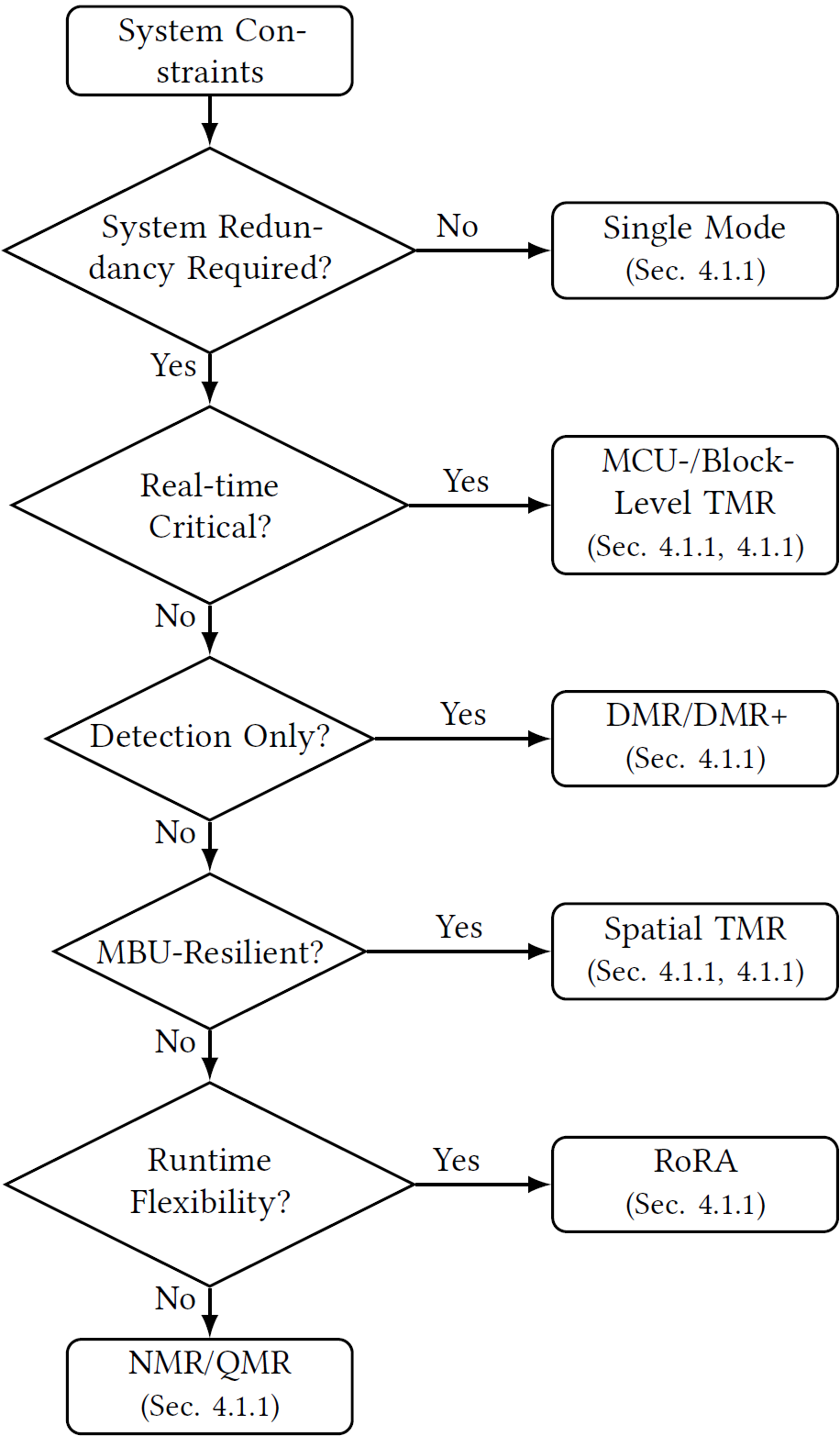}
    \caption{Decision flowchart for selecting spatial redundancy strategies.}
    \Description{A flowchart starting with 'System Constraints', flowing downwards through five decision diamonds. First, 'System Redundancy Required?' leads right to 'Single Mode' if No, or down if Yes. Second, 'Real-time Critical?' leads right to 'MCU-/Block-Level TMR' if Yes, or down if No. Third, 'Detection Only?' leads right to 'DMR/DMR+' if Yes, or down if No. Fourth, 'MBU-Resilient?' leads right to 'Spatial TMR' if Yes, or down if No. Fifth, 'Runtime Flexibility?' leads right to 'RoRA' if Yes, or down to a final block 'NMR/QMR' if No.}
    \label{fig:vertical_redundancy_flow}
\end{figure}

Spatial redundancy mitigates localized hardware faults by replicating functional units across physically distinct regions of an \gls{SoC} or full overarching system.
Redundancy models vary based on error type (transient, permanent), isolation capability, and area/power constraints~\cite{kastensmidt_optimal_2005, mitra_word-voter_2000, avizienis_n-version_1985}.
Figure~\ref{fig:vertical_redundancy_flow} outlines a decision framework. At a glance:
\begin{itemize}
    \item \textbf{No redundancy} is only viable in non-critical or testable environments.
\item \textbf{\gls{DMR}} is for detection-only systems where correction is offloaded or deferred.
\item \textbf{\gls{TMR}/\gls{NMR}} provide correction in-line and are essential for mission-critical systems.
\item \textbf{System-level \gls{TMR}} extends this protection beyond logic blocks to \gls{I/O}, power, and software domains~\cite{yeh_triple-triple_1996, baek_n-modular_2019}.
\end{itemize}

\paragraph{Decision Guidance (Fig.~\ref{fig:vertical_redundancy_flow})} 
The flowchart in Fig.~\ref{fig:vertical_redundancy_flow} supports the selection of appropriate spatial redundancy strategies based on system-level constraints, including real-time behavior, fault detection needs, \gls{MBU} tolerance, and runtime flexibility.

\paragraph{Single Module without Redundancy} \label{par:single_mode}
Before starting with the redundancy techniques, it is paramount to define the baseline case: a single-module system with no built-in redundancy.
In such configurations, the system operates with only one instance of each functional component.
Any hardware fault or transient error directly affects the system output, as there is no mechanism for fault masking or correction.
Fault detection, if present, relies entirely on external monitoring mechanisms such as watchdog timers, software assertions, or supervisory control logic.
This configuration offers no inherent fault tolerance. As Ganesan \emph{et al.} note in their study of distributed file systems, \textit{“a single file-system fault can cause catastrophic outcomes such as data loss, corruption, and unavailability”}~\cite{ganesan_redundancy_2017}.
Thus, the non-redundant case forms a critical baseline for evaluating the benefits of redundancy-based architectures.

\paragraph{Dual Modular Redundancy (DMR) and DMR Plus} \label{par:dmr}
This architecture employs two replicas and an XOR comparator to \emph{detect} mismatches, asserting an error flag whenever their outputs diverge.
In a DMR‑based AES S‑box, Taheri \emph{et al.} explain that \textit{“if the register’s contents are identical, no fault(s) is (are) detected. Otherwise the comparator… will be activated.”} and upon mismatch \textit{“the voter holds the previous value until the two replicas’ outputs become similar.”} \cite{taheri_dmr-based_2021}
Without a third replica, standard DMR cannot mask errors in-line. Consequently, any detected fault must be managed via external mechanisms, such as system resets or supervisory control. As an example of such a mechanism preventing error propagation, Taheri \emph{et al.} describe a DMR‑based AES S‑box where, upon detecting a mismatch, \textit{“the voter holds the previous value until the two replicas’ outputs become similar.”} \cite{taheri_dmr-based_2021}. While this halts the spread of corruption, it does not correct the underlying fault. 

Reviriego \emph{et al.} map each register pair in \gls{SRAM}-based \gls{FPGA}s to distinct configuration frames and use bitwise comparison logic to detect all single-bit upsets---yet, without a third replica, \gls{DMR} cannot mask errors in-line~\cite{reviriego_dmr_2016}.

Consequently, any detected fault in a DMR system must be recovered via external mechanisms—system resets, configuration scrubbing, or supervisory control—to restore correct operation once an error is flagged.
\\
Dual Modular Redundancy Plus (DMR+) augments plain \gls{DMR} with targeted recovery: when a parity mismatch is detected, the faulty register is overwritten by its golden copy while the other replica continues operation;
if no error is found, the first copy is used, otherwise the second copy is selected—ensuring correct output in either case.
Compared to full \gls{TMR}, this selective scrubbing reduces \gls{LUT} overhead by 50\%\ and flip‑flop overhead by 16.7\%\ (see Table 1 in set reference paper) \cite{reviriego_dmr_2016}, but—as with plain \gls{DMR}—it cannot tolerate permanent hardware failures without repeated scrubbing to restore to the correct state.

\paragraph{Triple Modular Redundancy (TMR)} \label{par:tmr}
This method triplicates each module and uses majority voting to mask faults.
The system functions correctly if either none or exactly one of the three modules fails.
The reliability of the TMR system, denoted \( R \), is then given by
\begin{equation}
    R = R_M^3 + 3R_M^2(1 - R_M) = 3R_M^2 - 2R_M^3,
\end{equation}
where \( R_M \) is the reliability of an individual module~\cite{lyons_use_1962}.

\paragraph{MCU‑Level TMR} \label{par:mcu_tmr}
This instantiates three identical processor cores that execute the same instruction stream in perfect lockstep, with a cycle‑by‑cycle majority voter sampling key architectural state (e.g. program counter and general‑purpose registers).
Baek \emph{et al.} \textit{“The TL-NMR framework determines the number of copies of each job…ensuring that every copied job completes its execution before its absolute deadline”} and demonstrate that \gls{TL-NMR} \textit{“maintains the schedulability, while significantly improving average system safety”} for transient fault rates up to $10^{-2}$~\cite{baek_n-modular_2019}.
By contrast, truly hardware‑based MCU‑TMR schemes necessitate sub‑nanosecond clock‑skew tolerances—since single‑event transients can last nearly 2ns, corrupting multiple cycles at frequencies above 500MHz—and require isolated power/ground domains, e.g.
\textit{“the use of an insulating substrate or an epitaxial layer…eliminates the formation of the \gls{SCR} circuit”} to suppress \gls{SEU}/\gls{SEL} crosstalk\cite{label_radiation_2005};
yet no public study has quantified the precise alignment margin or the minimum domain isolation distance needed to guarantee single‑event upset immunity.
Thus a deeper, public understanding of MCU triplication and its effects has to further be evaluated.

\paragraph{System‑Level TMR} \label{par:system_tmr}
This extends fault tolerance beyond individual cores by triplicating entire subsystems—including processors, I/O interfaces, power supplies—and employing hierarchical voting at multiple architectural layers.
Yeh’s “Triple‑Triple” architecture for the Boeing777 \textit{“uses triple redundancy for all hardware resources: computing system, airplane electrical power, hydraulic power and communication path”} and further notes that the Boeing777 primary flight computers \textit{“consist of three similar channels (of the same part number), and each channel contains three dissimilar computation lanes”} to mask faults at both hardware and software levels~\cite{yeh_triple-triple_1996}.
Despite this high‑level description, quantitative details—such as inter‑subsystem vote latency, the timing hierarchy between channel‑, lane‑, and bus‑level voting, or the minimum physical separation required to prevent common‑mode failures—remain unpublished, representing a clear gap in the open literature.
ISO26262 defines \textit{“hardware architectural metrics”} – the single‑point fault metric and the latent‑fault metric – and explicitly cites \textit{“Duplicated functional components (3.21)…for the purpose of increasing availability (3.7) or allowing fault (3.54) detection”} as an example of redundancy to meet \gls{ASIL} requirements~\cite{noauthor_iso_nodate}.
Yet, beyond these abstract metrics, \textit{“the technical mechanisms and solutions remain mainly open,”} and public descriptions of actual voter designs, component selections, or placement guidelines in automotive \gls{ECU}s are lacking~\cite{stoffel_-demand_2024}.
Stoffel \emph{et al.} further note that \textit{“one option is using Triple Modular Redundancy (TMR) which is tripling all components,effecting the cost, need of space and energy consumption”}, but do not detail the voter implementation, alignment tolerances, or isolation distances required for robust on‑demand TMR in production systems~\cite{stoffel_-demand_2024}.

In the railway domain, De Beck \emph{et al.} describe a programmable interlocking architecture based on \textit{“three interlocking calculators in charge of the safe functions”} that \textit{“periodically compare their opinions”} via a dedicated comparison processor, and automatically switch off any discordant unit, with dual optical‑fiber rings for lineside interface redundancy~\cite{de_beck_programmable_1987}.
However, quantitative details such as the vote‑cycle latency, macro‑cycle timing margins, or minimum physical separation of redundant channels are not specified in their public report, leaving a gap in guidelines for deployment.
Although Airbus A320 certification documents (e.g.\ \gls{EASA} CS-25) mandate Normal, Alternate and Direct flight-control laws with triple redundancy~\cite{noauthor_certification_nodate}, no open technical source details the voting algorithms or physical partitioning rules—an important omission that hampers academic replication and cross‐domain innovation.

\paragraph{MBU Mitigation via Placement} \label{par:mbu_mitigation}
As process geometries in both \gls{FPGA} and \gls{MCU} devices shrink to 28nm and below —have only recently entered mainstream production— the charge deposited by a single heavy‐ion strike can upset clusters of adjacent cells, leading to \gls{MBU}s that defeat some forms of spatial TMR layouts~\cite{moyer_legacy_2024, goldberg_how_2025}.
Quinn \emph{et al.} report that, on Xilinx Virtex devices, \gls{MBU}s account for 21\,\%\ of upsets at 150nm (\gls{LET} = 58.7MeV·cm²/mg) and rise to 59\,\%\ at 65nm (\gls{LET} = 68.3MeV·cm²/mg), with non‐normal incidence strikes producing up to 72\,\%\ \gls{MBU}s of four or more bits.
They emphasize that \textit{“placement and routing tools need to place the three separate domains far enough apart to not be affected by MBUs”} \cite{quinn_domain_2007}.
LaBel and Cohn survey circuit‑layout mitigations, noting that \textit{“word bit physical location staggering and critical node interleaving”} can preclude a single ion from impacting multiple bits of the same word, and that the use of \gls{SOI} substrates yields an order‑of‑magnitude reduction in \gls{SEU}/\gls{MBU} rates by shrinking charge‑collection volumes \cite{label_radiation_2005}.
Despite these isolated strategies, no public study has yet derived general, quantitative guidelines relating process node, inter‑domain spacing, and \gls{MBU} probability.
This absence of continuous placement times node scaling laws leaves designers without the means to predict and mitigate \gls{MBU} vulnerability across today’s and future sub‑28nm technologies.

\paragraph{Detailed Hardware Replication} \label{par:block_tmr}
\begin{itemize}
    \item \textbf{Logic‑Level TMR} replicates every logic gate and voter within the design. While this achieves the finest‑grained fault masking, it incurs severe area and performance penalties: in Gaisler’s \gls{FTMR} implementation, combinatorial logic overhead ranged from factor 4.5 to factor 7.5 (i.e.\ 350–650\,\%\ increase) and overall SLICE usage grew by a factor of 5.25–6.25, with a corresponding overall throughput drop of approximately 50\,\%\ \cite{habinc_sandi_functional_2002}.

    \item \textbf{Register‑Level TMR} applies TMR only to flip‑flops and adds self‑checking voters at each register output. Afzaal and Lee’s on‑chip “self‑checking voting circuit” uses redundant voter copies and a two‑stage consensus comparator. They proof functionality by fault injection to be \textit{“averaging the reliability values...for p = 1 (perfect switching case) versus p = 0 (unprotected simplex voter), we roughly obtain a 26\,\%\ improvement in reliability”} \cite{afzaal_self-checking_2018}.

    \item \textbf{Block‑Level TMR} groups entire IP blocks (e.g.\ ALUs, memory controllers) under a single voter. Mitra and McCluskey define the probability $s$ as
    \[
        s = \Pr\left(\substack{
            \text{word‑voter signals error at least once before}\\
            \text{an incorrect output, given multiple faults}
        }\right)
    \]
    and demonstrate via simulation (100,000 runs on multiple different logic circuits) that a higher probability $s$ yields orders‑of‑magnitude data‑integrity gains (e.g.\ $s=0.5$ vs.\ $s=0.99$). The word‑voter’s hardware cost is exactly one extra 2‑input gate plus three XNORs per output bit—equivalently $(6n+2)$ two‑input gates and $3n$ XNORs for an $n$‑bit block—versus $5n$ gates for bit‑wise voting, with logic blocks \cite{mitra_word-voter_2000}.
\end{itemize}

\paragraph{Distributed TMR} \label{sec:distributed_tmr}
This category further refines spatial redundancy by varying how the redundant modules (replicas) are physically placed, the extent to which the system is triplicated, and whether this redundancy is fixed at design time or dynamically managed:

\begin{description}
\item[\textbf{Full (Global) TMR}]
    also called \emph{Spatial TMR}, here each replica and its voter are placed in completely distinct clock, power, and routing domains, often via physical separation at the board or die level.
Kastensmidt \emph{et al.} demonstrate that such isolation reduces \gls{SEU}-induced failures from 5,321 to 727 upsets, roughly a factor 7.3 improvement, albeit with increased latency and board-level complexity \cite{kastensmidt_optimal_2005}.
\item[\textbf{Partial TMR}]
    triplicates only a subset of the design, typically control paths or configuration-sensitive registers.
Pratt \emph{et al.} show that applying \gls{TMR} selectively to feedback structures increases slice usage by only 40\,\% (5,746 $\rightarrow$ 8,036 slices) compared to over 90\,\% for full \gls{TMR}, effectively reducing mitigation overhead while preserving protection against persistent \gls{SEU}s \cite{pratt_improving_2006}.
Sanchez-Clemente \emph{et al.} extend this by injecting approximate logic into non-critical paths to maintain acceptable output fidelity under \gls{SEU} exposure \cite{sanchez-clemente_partial_2016}.
\item[\textbf{Localized TMR}]
    groups replicas within adjacent \gls{FPGA} frames or logic tiles to minimize routing overhead and skew.
Sterpone and Violante (2005) investigated \gls{TMR} robustness on a Xilinx Spartan by comparing 'no constraints', 'minimal-area', and 'safe-area' floorplanning strategies.
While large circuits under 'no constraints' showed up to 13.18\,\%\ "Wrong Answers" from injected \gls{SEU}s, 'safe-area' placement drastically reduced this to $\approx0.23\%$ for a Mul4 circuit, underscoring the importance of isolated module placement.
Building on this, Sterpone \emph{et al.} (2011) extended the analysis to Multi-Cell Upsets (\gls{McellU}s), employing laser testing and a STAR-\gls{McellU} tool to incorporate precise physical layout information.
Their findings revealed that \gls{McellU}s can significantly degrade \gls{TMR} effectiveness by inducing multiple, simultaneous errors across redundant modules, with 2-bit \gls{McellU}s corrupting \gls{TMR}-hardened circuits 2.6 orders of magnitude more than \gls{SBU}s, emphasizing the critical need for layout-aware design \cite{sterpone_analysis_2005, sterpone_layout-aware_2011}.
\item[\textbf{Static TMR}]
    fixes replica placement at design time with no in-field reconfiguration.
This relies on robust synthesis-time partitioning—guard-banded routing, over-provisioned timing margins, and conservative floorplanning—to maintain isolation throughout the device’s lifetime. Placement strategies in modern FPGA architectures, including multi-die designs, further highlight the importance of interconnect organization and spatial partitioning \cite{wittlichASurveyOfFPGAPlacement}.
Habinc and Sandi’s Functional TMR (\gls{FTMR}) templates demonstrate significant overheads for gate-level triplication.
For a demonstration application targeting a Xilinx Virtex XCV1000-6 device, they report a resource increase by a factor between 4.5 and 7.5 in terms of slice usage, coupled with an observed performance decrease of about 50\,\%\ \cite{habinc_sandi_functional_2002}.
The document does not explicitly recommend an additional X\,\%\ clock-rate margin or redundant decoupling capacitors to tolerate process drift and aging over a fixed time period e.g.
10-year mission profile.

\item[\textbf{Reliability-oriented Place and Route Algorithm (RoRA)}] \label{par:rora}
    is a static, reliability-oriented placement and routing tool designed to mitigate \gls{SEU} effects in the configuration memory of \gls{SRAM}-based \gls{FPGA}s. Instead of using dynamic reconfiguration or dormant replicas, \gls{RoRA} statically partitions the \gls{FPGA} resources into distinct, physically isolated sets during the design phase. This ensures that a single configuration bit upset affecting a routing matrix (switch box) cannot bridge or disconnect wires across multiple redundant channels simultaneously, preventing specific "short" or "open" faults that would defeat standard \gls{TMR}. Sterpone \emph{et al.} implemented \gls{RoRA} on a Xilinx Spartan XC2S200, reporting that it does not introduce any area overhead compared to traditional \gls{TMR} (which remains about 3 times larger than the plain circuit). However, the dependability-oriented routing constraints resulted in circuits that were 22\,\%\ slower on average than their standard \gls{TMR} counterparts. Despite this performance penalty, \gls{RoRA} drastically improved resilience against configuration memory upsets, increasing \gls{SEU} tolerance by up to 85 times for the tested benchmarks \cite{sterpone_rora_2005}.
    
    Complementing this static approach with dynamic repair, Shinohara and Watanabe extended reliability concepts into a Double or Triple Module Redundancy method for dynamically reconfigurable devices. Their method utilizes spare frames to reduce hardware cost to approximately 66\,\%\ of traditional \gls{TMR} when the \gls{SEU} occurrence probability is near zero, effectively trading static protection for on-demand reconfiguration \cite{shinohara_double_2008}. To support this dynamic approach, they highlight that Optically Reconfigurable Gate Arrays (\gls{ORGA}s) used in their experiments inherently possess high defect tolerance and rapid recovery capabilities, making them highly suitable for such repeated reconfiguration cycles \cite{shinohara_double_2008}.
\end{description} 

\paragraph{Scaling to NMR and QMR} \label{par:nmr}
Generalized as \gls{NMR}, this uses $N$ odd replicas to tolerate up to $\left\lfloor \frac{N-1}{2} \right\rfloor$ faults.
The concept of N-Modular Redundancy, particularly the "N-version approach," has been a foundational method for achieving fault tolerance by utilizing multiple independently developed versions of a module to perform the same task, with a voter determining the correct output \cite{avizienis_n-version_1985}.
While higher-order redundancy schemes like \gls{NMR} aim for increased fault tolerance, they typically come at the cost of significant area and power overhead, and increased voter complexity.
For instance, Amiri and Přenosil discuss methods to significantly improve the reliability of \gls{NMR} systems, highlighting the inherent trade-offs involved in achieving higher fault tolerance \cite{amiri_significant_2015}.
These methods focus on enhancing the overall system reliability, implying that without such improvements, the benefits of scaling to higher (N) might be mitigated by practical limitations.
%--------------------------------------------------------------------
%--------------------------------------------------------------------
%--------------------------------------------------------------------

\subsubsection{Temporal Redundancy}\label{sec:temporal_redundancy}

\begin{figure}[htbp]
    \centering
    \includegraphics[width=0.55\linewidth]{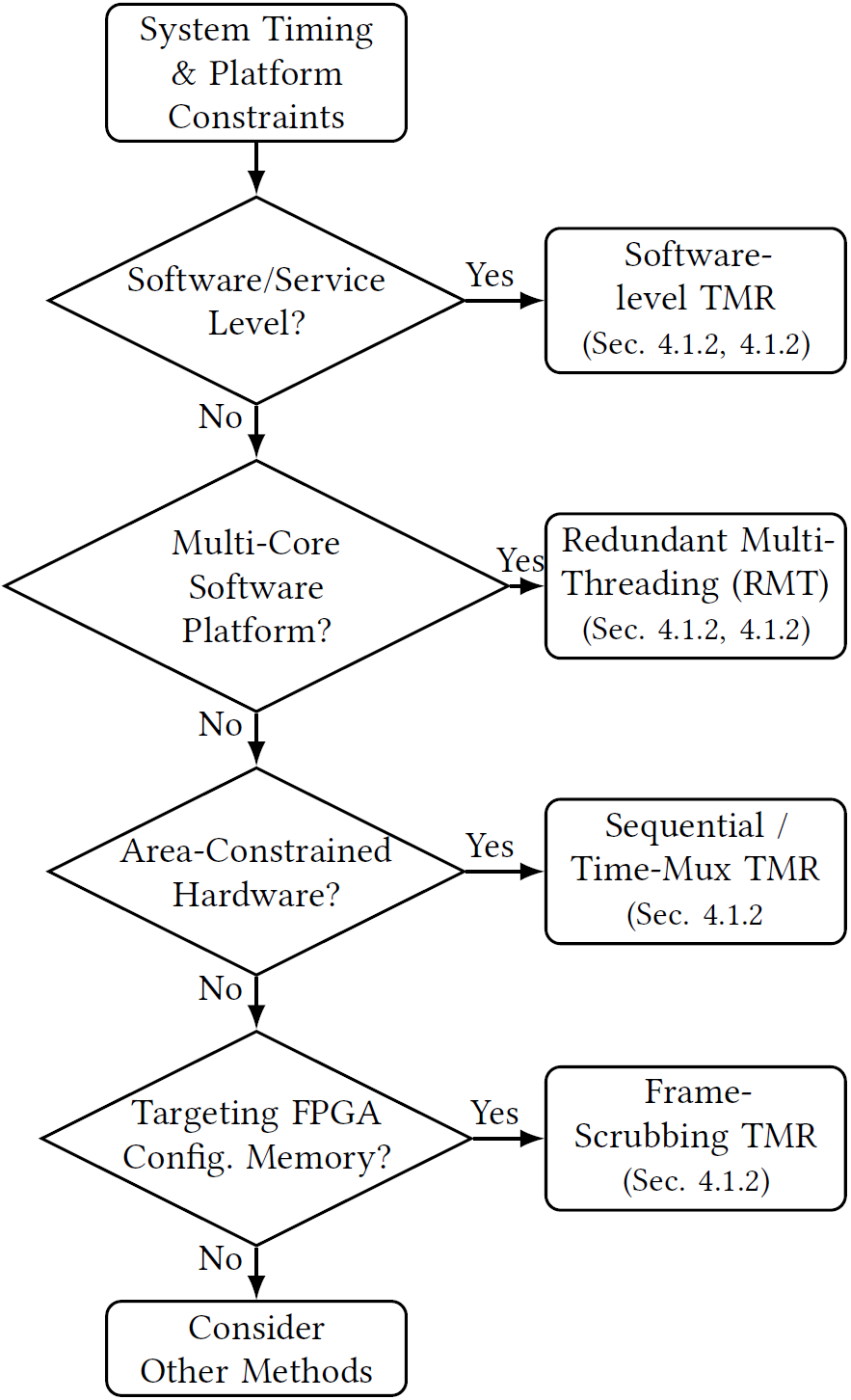}
    \caption{Navigation tree for selecting temporal redundancy techniques.\label{fig:temporal_flow}}
    \Description{A flowchart starting with 'System Timing and Platform Constraints', proceeding downward through four decision diamonds. 'Software/Service Level?' leads right to 'Software-level TMR' if Yes, or down if No. 'Multi-Core Software Platform?' leads right to 'Redundant Multi-Threading (RMT)' if Yes, or down if No. 'Area-Constrained Hardware?' leads right to 'Sequential / Time-Mux TMR' if Yes, or down if No. 'Targeting FPGA Config. Memory?' leads right to 'Frame-Scrubbing TMR' if Yes, or down to a final block 'Consider Other Methods' if No.}
\end{figure}

Temporal redundancy tolerates transient faults through \emph{time-based
replication}: the same hardware is exercised in multiple time windows
and the results are compared or refreshed.
When board area or mass
excludes full spatial triplication, temporal methods provide an
economical path to increased reliability.
Compared with spatial
\gls{TMR}, temporal techniques must pay in latency or throughput;
therefore,
the first design question is always \emph{“How much slack exists in my
real-time budget?”} Figure~\ref{fig:temporal_flow} answers that
question in a decision tree.

\paragraph{Decision Guidance (Fig.~\ref{fig:temporal_flow})}
Figure~\ref{fig:temporal_flow} presents a high-level 
decision tree for selecting among temporal redundancy techniques based on the system's timing paradigm, target platform, and primary constraints.

\paragraph{Dynamic TMR}  \label{par:dynamic}
Keeps replicas, \emph{virtual} until a vote is needed \cite{anwer_dynamic_2020}.

\paragraph{Frame-Scrubbing TMR} \label{par:scrubbing}
Pratt \emph{et al.} presented a fault-injection framework for assessing reliability improvements of scrubbing and \gls{TMR} on Xilinx Virtex-II \gls{FPGA}s \cite{pratt_improving_2006}.
Their "Partial TMR" approach selectively applies mitigation to critical circuit structures, aiming to reduce costs while improving reliability.
Configuration scrubbing, integral to their method, continuously reads and evaluates \gls{FPGA} configuration memory to detect and repair radiation-induced faults by reloading correct data without halting operation.
This technique exclusively targets configuration memory faults, not data-path or register-level upsets.\cite{pratt_improving_2006}.
Herrera-Alzu and López-Vallejo extended this with a self-reference scrubber for inter-\gls{FPGA} \gls{TMR} systems on Xilinx Virtex \gls{FPGA}s~\cite{wittlichASurveyOfFPGAPlacement}, including the radiation-hardened Virtex-5QV \cite{herrera-alzu_self-reference_2011}.
Their scrubber uses peer frame comparison for fast \gls{SEU} and \gls{MBU} detection and correction, eliminating the need for a golden configuration memory.
\gls{SRAM}-based \gls{FPGA}s like the Virtex-4QV and Virtex-5QV are susceptible to radiation, necessitating configuration memory scrubbing for space applications \cite{herrera-alzu_self-reference_2011}.
The Virtex-5QV, a "rad-hard-by-design" device, offers high resistance to \gls{SEU}s, total immunity to \gls{SEL}s (less then 100 MeV per mg-cm²), and a \gls{TID} tolerance exceeding 1 Mrad(Si) \cite{xilinx_inc_radiation-hardened_2018}.
It features \gls{SEU}-hardened configuration memory and control logic, with very low orbital upset rates.
For Virtex-4 and later \gls{FPGA}s, a configurable frame is 41 words of 32 bits, supporting runtime scrubbing via dynamic partial reconfiguration\cite{herrera-alzu_self-reference_2011}.
This scrubber also allows dynamic scrub rate adjustment and fast \gls{SEFI} detection, applicable to Xilinx \gls{FPGA}s from Virtex-4 onwards.

\paragraph{Sequential or Temporal Redundancy} \label{par:time_mux}
Can be implemented at different architectural levels, either by time-sharing hardware or by re-executing software tasks.
At the hardware level, Time-Multiplexed TMR mitigates \gls{SEU}s using a single functional module and voter that are time-shared, as presented by She and McElvain.
The hardware sequentially processes three redundant data threads, and a majority vote on the results masks any single transient fault \cite{she_time_2009}.
This approach drastically reduces area and power overhead compared to spatial \gls{TMR} at the cost of increased latency, making it ideal for resource-constrained, delay-insensitive systems.
At the system level, Chen \emph{et al.} investigate a purely temporal fault-tolerance scheme via task re-execution, or Sequential TMR, in multi-core systems.
Their work on redundant multithreading enables the re-execution of tasks on either the same or different cores, with a majority-voting mechanism ensuring correctness.
Since transient faults are unlikely to persist across re-executions, this method significantly improves reliability.
The authors propose an optimal, reliability-aware task-to-core mapping scheme that can be managed by a compiler to ensure both performance and fault resilience, making it well-suited for applications where soft error resilience is critical and timing constraints are moderately flexible \cite{chen_task_2016}.
Both techniques thus apply the same core principle of temporal redundancy but target different layers of the system stack.
\begin{description}
    \item[Asynchronous TMR] \label{par:async}
    Enhances robustness by removing the global clock as a single point of failure, instead relying on self-timed logic and local handshaking protocols.
In such systems, fault tolerance is achieved not through clocked re-execution, but through delay-insensitive temporal coordination.
Balasubramanian \emph{et al.} explore this paradigm through the design of \gls{QDI} majority voters, where the voter's operation is event-driven;
it waits to receive valid data from all three modules, signaled via a handshake protocol, before computing the majority output \cite{balasubramanian_area_2019}.
This inherent delay insensitivity ensures the system can tolerate timing variations or stalls in one module without producing an incorrect result.
While the module replication is spatial, the fault masking relies critically on this self-timed temporal behavior.
A more detailed discussion of this voter implementation can be found in Section~\ref{subsec:Optimization-Oriented_Voters}~(B).
\item[Synchronous Temporal Variants] 
    Apply deterministic, clock-driven execution to recover from transient faults without the area overhead of full spatial triplication.
These mechanisms are well-suited for systems with tight timing constraints where the core principle is to leverage time---through re-computation, periodic checks, or enhanced synchronization---to ensure data integrity within a clocked framework.
Key strategies are often realized through specialized voter designs. For instance, the temporal consistency of data can be enforced by voting on entire data words rather than individual bits, a technique detailed in Section~\ref{subsec:basic_voters}.
Another approach involves actively refreshing the state of sequential elements using periodic, majority-based overwrites, as described in Section~\ref{subsec:Adaptive_Dynamic_Voters}~(D).
Furthermore, the reliability of the voter itself can be enhanced with internal self-checking mechanisms, which are covered in Section~\ref{subsec:feedback-based_Voter}.
Finally, maintaining synchronization is paramount, especially in dynamic systems. Pilotto \emph{et al.} address the challenge of re-synchronizing a TMR module after it has been restored via dynamic partial reconfiguration.
Their method uses pre-defined check-point states, allowing the recovered module to be held until the other two correct modules reach the same state, at which point all three resume lockstep operation without system downtime \cite{pilotto_synchronizing_2008}.
These varied approaches demonstrate that synchronous temporal redundancy offers a rich set of deterministic, clock-driven solutions for fault tolerance.
\end{description}

\paragraph{Redundant Multi-Threading (RMT)} \label{par:rmt}
Also known as \gls{TMT}, achieves fault tolerance by executing redundant copies of an application's threads.
This software-level approach can be implemented in various ways across different system architectures and layers.
On general-purpose multi-core processors, \gls{RMT} can be enabled by an operating system service, as demonstrated by Döbel and Härtig.
Their system provides replication for unmodified multithreaded binary applications by transparently enforcing a deterministic execution order for synchronization operations, which allows for correct state comparison and majority voting \cite{dobel_can_2014}.
For systems without an \gls{OS}, Serrano-Cases \emph{et al.} implement \gls{RMT} on a bare-metal ARM platform.
Using static barriers and custom macros for synchronization, their approach detected and recovered from all observed \gls{SDC}s under neutron irradiation, though it also increased the system's susceptibility to hangs due to thread desynchronization \cite{serrano-cases_bare-metal_2023}.
The \gls{RMT} concept has also been adapted to specialized architectures.
On \gls{GPU}s, Milluzzi and George use persistent threads and \gls{CUDA} Streams to replicate critical kernels across streaming multiprocessors.
This approach mitigates faults and reduces the load on the non-deterministic hardware scheduler, achieving a factor of 1.58 speedup over an equivalent \gls{TMR} implementation running on the \gls{CPU} cores \cite{milluzzi_exploration_2017}.
For \gls{RISC-V} cores, Barbirotta \emph{et al.} present a dynamic \gls{RMT} scheme that operates in an efficient \gls{DMR} mode for fault detection and activates a third, \gls{DTMR} thread for \gls{TMR}-style recovery only when a fault occurs.
This dynamic approach achieves a 33\,\%\ performance gain and a \gls{LUT} reduction in 25.3\,\%\ over a continuous TMR implementation \cite{barbirotta_evaluation_2022}.
Finally, managing these redundant threads efficiently is critical. 

Chen \emph{et al.} address the challenge of mapping redundant threads onto multi-core systems with heterogeneous performance characteristics.
They formalize the task mapping as a constraint optimization problem, allowing a system to decide whether tasks should run in a redundant or non-redundant mode to maximize overall reliability while satisfying timing constraints.
Compared to state-of-the-art greedy mapping strategies, their approach achieves up to 80\,\%\ reliability improvement, with an average of 20\,\%\ across different, tested, chip frequency variation scenarios \cite{chen_task_2016}.

\paragraph{Software-level TMR}  \label{par:swtmr}
Replicates function calls or threads and votes on their results, offering a cost-effective redundancy option.
The foundational concept is Avizienis' \textit{N-Version Programming}, which proposed using independently developed software versions to tolerate software design faults through diversity~\cite{avizienis_n-version_1985}.
Modern approaches adapt this principle to tolerate transient hardware faults in \gls{COTS} hardware.
These techniques are applied across various system structures. For distributed embedded systems, Janson \emph{et al.}'s STROBES framework uses \gls{CRC}-based checks to achieve robust fault detection across loosely synchronized nodes without requiring special hardware~\cite{janson_software-level_2018}.
For bare-metal multicore systems, Serrano-Cases \emph{et al.} implement \gls{RMT} without an \gls{OS}, demonstrating full recovery from all observed \gls{SDC}s under neutron irradiation, though with an increased susceptibility to system hangs~\cite{serrano-cases_bare-metal_2023}.
To support unmodified multithreaded applications, Döbel and Härtig developed an \gls{OS} service that enforces deterministic execution of synchronization points, allowing for correct state comparison and voting~\cite{dobel_can_2014}.
Further system-level optimizations, such as the reliability-driven task mapping for heterogeneous cores proposed by Chen \emph{et al.}, aim to maximize reliability while respecting performance constraints~\cite{chen_task_2016}.
\subsubsection{Mixed Redundancy} \label{sec:mixed_redundnacy}
Integrates spatial and temporal fault-tolerance with adaptable or partial mechanisms.
This chapter presents techniques such as adaptive \gls{TMR}, reconfigurable designs, and hybrid schemes that adjust fault response dynamically.
Their flexible nature and architectural diversity warrant a separate classification, due to its interlinked nature and emerging techniques.
\begin{figure}[htbp]
    \centering
    \includegraphics[width=0.55\linewidth]{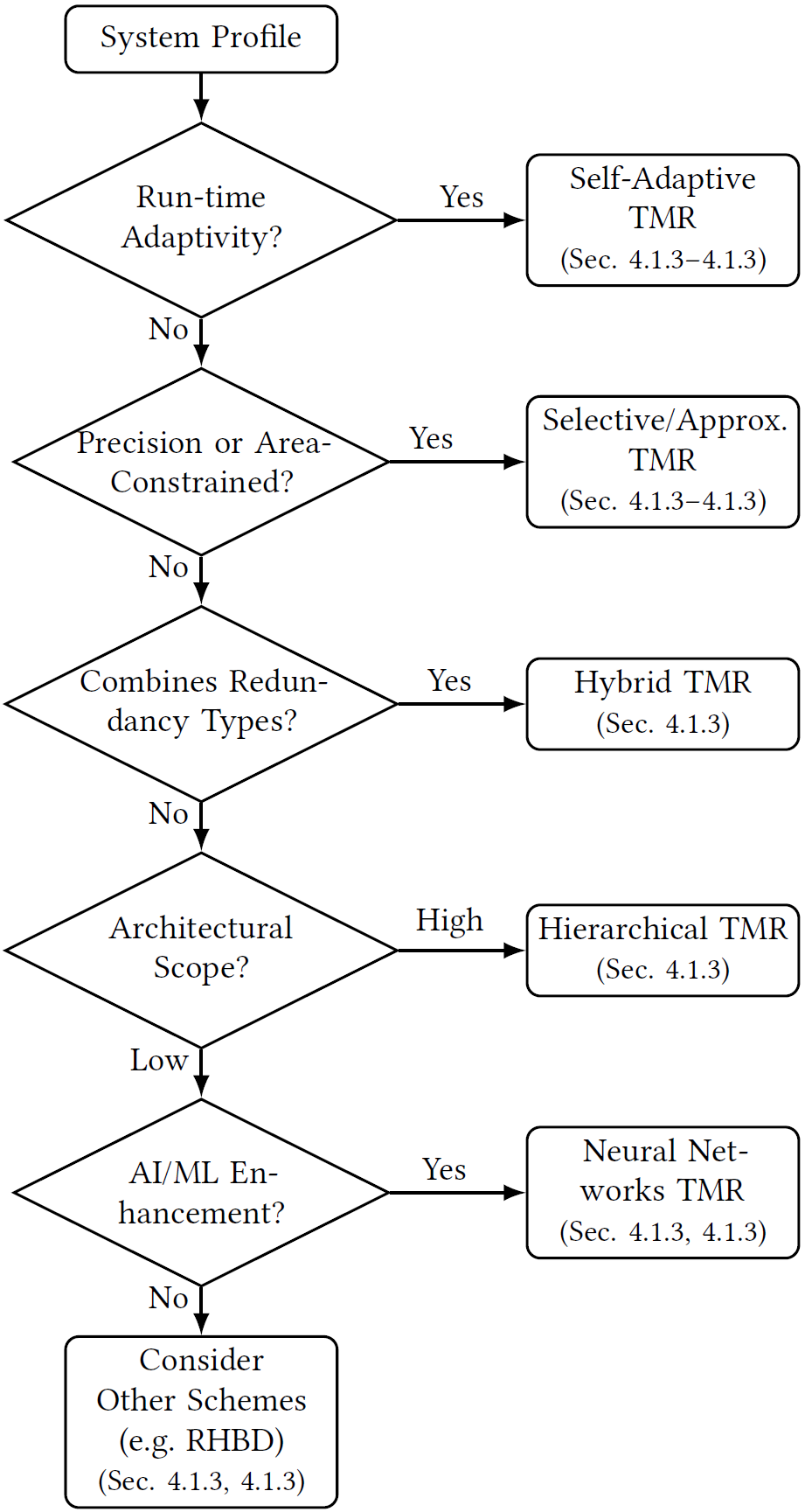}
    \caption{Decision tree for selecting mixed redundancy strategies.}
    \Description{A flowchart starting with 'System Profile', flowing downwards through five decision diamonds. 1: 'Run-time Adaptivity?' branches right to 'Self-Adaptive TMR' on Yes, down on No. 2: 'Precision or Area-Constrained?' branches right to 'Selective/Approx. TMR' on Yes, down on No. 3: 'Combines Redundancy Types?' branches right to 'Hybrid TMR' on Yes, down on No. 4: 'Architectural Scope?' branches right to 'Hierarchical TMR' on High, down on Low. 5: 'AI/ML Enhancement?' branches right to 'Neural Networks TMR' on Yes, down to a final block 'Consider Other Schemes (e.g. RHBD)' on No.}
    \label{fig:mixed_redundancy_flow}
\end{figure}

\paragraph{Decision Guidance (Fig.~\ref{fig:mixed_redundancy_flow})}
Provides a logic route for identifying suitable mixed redundancy techniques based on key system characteristics.

\paragraph{Self-Adaptive TMR} \label{par:self_adaptive}
Encompasses techniques where the redundancy configuration adapts dynamically to environmental conditions or operational needs.
Reconfigurable \gls{TMR} allows swapping in cold spares when faults are detected~\cite{sterpone_rora_2005}, while fault-aware schemes monitor error patterns to adapt redundancy as needed~\cite{barbirotta_evaluation_2022}.
Selective \gls{TMR} variants (e.g. \gls{ATMR}~\cite{balasubramanian_fault-tolerant_2023}) reduce power by triplicating only critical or approximate logic paths.
These strategies emphasize flexibility and efficiency.

\paragraph{Reconfigurable TMR} \label{par:reconfigurable}
Leverages dynamic partial reconfiguration to replace faulty modules without interrupting system operation, using pre-placed or logically mapped spares in unutilized \gls{FPGA} regions.
A key challenge in this paradigm is managing the on-chip communication required to trigger the \gls{FPGA}'s reconfiguration process.
Agiakatsikas \emph{et al.}~\cite{agiakatsikas_reconfiguration_2016} address this by introducing Reconfiguration Control Networks (\gls{RCN}s). An \gls{RCN} acts as a dedicated on-chip routing infrastructure: when a local voter detects a fault, it sends a reconfiguration request through the \gls{RCN} to a central hardware controller. This controller then executes the actual repair by writing new configuration data through the \gls{FPGA}'s Internal Configuration Access Port (\gls{ICAP}).
Their evaluation on a Xilinx Artix-7 platform compared different \gls{RCN} routing topologies and revealed a critical trade-off: a minimal, \gls{ICAP}-centric bus topology minimized resource usage (requiring a factor of 2.7 fewer essential bits compared to other topologies), but its request routing latency was over a factor of 175 greater than employing a dedicated token-ring network topology.
To address the \gls{RCN}'s own susceptibility to faults, the authors recommend its full triplication and internal error correction.

An alternative approach, \gls{TMRSB}, was proposed by Zhang \emph{et al.}~\cite{zhang_triple_2006}.
This model enhances standard \gls{TMR} by associating each replicated module with a dynamic pool of standby configurations.
Upon fault detection, a faulty module is remapped to a logically equivalent spare from its pool.
This provides greater flexibility than single-spare systems and explicitly models the reliability of the switching mechanism.
Reliability simulations using BlockSim demonstrated that \gls{TMRSB} surpasses both classic \gls{TMR} and cold-standby systems during the initial phases of a mission.
However, the reliability benefits were found to saturate beyond four standby configurations per module, an effect attributed to the increasing overhead and failure probability of the switching logic.

Further research has focused on refining reconfigurable \gls{TMR} through more localized repair strategies.
For instance, Shinohara \emph{et al.}~\cite{shinohara_double_2008} proposed a method of partial re-triplication triggered by localized error diagnosis to reduce reconfiguration overhead (see \ref{sec:distributed_tmr}: Reliability-oriented Place and Route Algorithm).
Similarly, Veljković \emph{et al.}~\cite{veljkovic_adaptive_2015} demonstrated how hardening interface-level logic allows for the isolation of failures, enabling the reconfiguration of only the specific logic tiles that are affected.
These techniques effectively use reconfigurability as a form of temporal redundancy layered on top of spatial \gls{TMR}, enabling flexible resilience trade-offs.
However, their practical application faces significant hurdles. The efficacy of these methods relies on robust, high-speed coordination and state synchronization between modules and voters, a challenge that is not always fully addressed.
Furthermore, as highlighted in a survey by Vipin and Fahmy~\cite{vipin_fpga_2018}, a major bottleneck remains the lack of high-level reconfiguration \gls{API}s and unified toolchains for commercial \gls{FPGA} development flows like Vivado, which complicates the integration of such dynamic fault-tolerance schemes.

\paragraph{Fault-Aware TMR} \label{par:fault_aware}
Is a design-time technique that adapts an implementation to known manufacturing defects or wear-out faults.
Instead of relying on runtime reconfiguration, it uses a pre-characterized map of faulty resources to guide the synthesis and place-and-route process, ensuring that logic is not mapped to unreliable physical regions of the \gls{FPGA}.
A prominent example is the fault-aware toolchain developed by Gupte \emph{et al.}~\cite{gupte_fault-aware_2015}, which integrates defect maps into a standard \gls{FPGA} flow.
By using vendor-provided `PROHIBIT` constraints, the toolchain is instructed to avoid specific faulty logic slices during placement.
This preemptive approach allows the place-and-route tools to find an optimal implementation using only reliable resources.
Their findings show that this method can tolerate up to 30\,\%\ faulty logic resources with a performance degradation of less than 10\%, all without requiring additional hardware or runtime monitoring.

\paragraph{Selective and Adaptive Redundancy} \label{par:selective}
These techniques balance resilience and resource cost by hardening only critical components at design time or by choosing a fault tolerance strategy dynamically at runtime.
Using Selective Replication, as a static approach applying \gls{TMR} only to the most vulnerable modules, identified via pre-deployment analysis.
Samudrala \emph{et al.}~\cite{samudrala_selective_2004} use signal probability analysis to find sensitive subcircuits.
Bertoa \emph{et al.}~\cite{bertoa_fault-tolerant_2023} apply this to neural network accelerators by hardening only high-impact computational channels, achieving significant protection with reduced hardware overhead.

\paragraph{Adaptive Fault-Tolerant Re-Execution} \label{par:aftr}
Strategically re-executes computations when fault indicators arise—merging temporal redundancy with dynamic system state awareness.
In multicore and distributed systems, this method enables on-demand redundancy allocation, preserving slack for critical paths while still addressing fault coverage.
Gonzalez \emph{et al.} \cite{gonzalez_adaptive_1997} propose a reflective runtime scheduler that dynamically selects between \gls{TMR}, \gls{PB}, and \gls{PE} strategies based on task criticality and available resources.
Their Spring kernel framework supports fallback options (e.g. \textit{\gls{AFT}-3}) and delayed strategy commitment, leading to \(10\text{--}20\,\%\) better task acceptance under overload, without lowering critical path reliability.
Bolchini \emph{et al.}\cite{bolchini_self-adaptive_2013} extend this paradigm to many-core embedded systems with fault-aware task replication.
Their scheduler exploits health metrics of individual cores to reroute failing threads.
This reduces overall resource consumption compared to static triplication while improving coverage under aging or radiation-induced degradation.
The approach can also quarantine recovery threads, as in \textit{hybrid \gls{SCP}-\gls{TMR}}~\cite{gonzalez_adaptive_1997}, to defer resource-intensive voting until deemed necessary.
These works highlight that \gls{AFTR} excels in dynamic systems—especially those with bursty task arrivals or intermittent fault behavior, instead of proactive full redundancy.
It offers \emph{opportunistic resilience}, which adapts protection strength to run-time conditions.

\paragraph*{Approximate TMR (ATMR) and Reliability Frameworks} \label{par:atmr}
\gls{ATMR} is a class of techniques that leverages approximate computing to reduce the area, power, and latency overheads of traditional \gls{TMR}, particularly in error-tolerant applications like signal processing. As comprehensive surveys by Ammes \emph{et al.}~\cite{ammes_survey_2024} and Arifeen \emph{et al.}~\cite{arifeen_approximate_2020} explain, \gls{ATMR} designs trade a degree of fault coverage for resource efficiency, facing challenges such as identifying input vulnerabilities and ensuring scalability~\cite{arifeen_input_2018}. Within this paradigm, strategies generally focus on protecting critical \gls{MSB}s while relaxing constraints on \gls{LSB}s. While the exact hardware implementation of this relaxation varies, both of the following approaches are commonly benchmarked using image processing tasks to objectively quantify the trade-off between perceptual quality and resource savings:

\begin{itemize}
    \item \textbf{Approximate Logic Integration (e.g., FAC):} Balasubramanian and Maskell~\cite{balasubramanian_fault-tolerant_2023} introduced \gls{FAC}, which partitions functional blocks into \gls{MSB} and \gls{LSB} domains. It applies precise triplication only to the critical \gls{MSB} parts while implementing the \gls{LSB}s using approximate logic circuits. Demonstrated on an \gls{FFT} image processing pipeline, this hybrid approach achieved a 19.5\% area reduction, 15.3\% less delay, and a 24.7\% power decrease compared to full \gls{TMR}, while maintaining high perceptual quality (\gls{PSNR} > 30 dB and \gls{SSIM} $\approx 1$).
    
    \item \textbf{Dynamic Truncation Frameworks (e.g., X-Rel):} Instead of replacing logic with approximate circuits, Vafaei \emph{et al.}~\cite{vafaei_x-rel_2023} introduce \gls{X-Rel}, an approximate reliability framework that adapts fidelity based on user-defined quality constraints. \gls{X-Rel} utilizes an \gls{ILP} formulation to determine the optimal number of \gls{LSB}s to completely truncate in each module. It jointly optimizes the system by reducing the voter to only compare the remaining \gls{MSB}s. Similar to \gls{FAC}, this co-optimization is validated via image quality metrics, achieving up to a 3.39x improvement in \gls{EDAP} and up to 114\% higher image quality (\gls{MSSIM}) compared to exact \gls{TMR} under error conditions.
\end{itemize}

Related work in the \gls{ATMR} domain has also focused on improving the reliability of the system's core decision logic, with fault-tolerant voter designs that integrate both error detection and recovery stages~\cite{arifeen_fault_2019}.

\paragraph{Hybrid TMR} \label{par:hybrid}
Unifies spatial and temporal redundancy to address multiple fault classes (for further insights see chapter ~\ref{sec:spatial_redundncy} and ~\ref{sec:temporal_redundancy}) while optimizing for cost and performance, a design philosophy effective in systems where fault rates or resource availability may change dynamically.
One approach combines spatial and temporal redundancy. An early example by Chu and Wah~\cite{chu_fault_1990} used multiple neurons (spatial), recomputation (temporal), and coding to protect neural.
A more recent variant, often called Dynamic or \gls{TTMR}, operates in an efficient dual-module mode and only invokes a third, time-redundant execution upon detecting a fault.
This "\gls{DMR} first, \gls{TMR} on-demand" strategy significantly reduces overhead and has been demonstrated in both general reconfigurable systems~\cite{czajkowski_seu_2006} and modern \gls{RISC-V} multi-threaded cores~\cite{barbirotta_evaluation_2022}.
Other hybrid approaches focus on repair, resource substitution, or diversity.
The \gls{TMRSB} model, proposed by Zhang \emph{et al.}~\cite{zhang_triple_2006}, enhances a standard \gls{TMR} system with a pool of standby configurations for dynamic repair.
A prominent hybrid strategy is Hardware-Software Co-TMR, which combines hardware speed with software flexibility.
The concept involves replicating a function across both a hardware accelerator (e.g. \gls{FPGA}) and a software platform (e.g. a core on the same \gls{SoC}) to provide fault tolerance through diversity~\cite{bingbing_xia_fault-tolerant_2010}.
Platforms like the Zynq \gls{SoC} are well-suited for such co-designs.
For example, Lede \emph{et al.}~\cite{lede_hardware_2023} demonstrate a HW/SW co-design for a real-time impact detection system, partitioning sensor preprocessing to an \gls{FPGA} pipeline and neural network classification to the ARM cores.
While their work focuses on performance and resource optimization—achieving under 52.84 ms latency with 18.91\,\%\ \gls{LUT} utilization—it illustrates the type of heterogeneous partitioning that is a prerequisite for Co-TMR.
However, true Co-TMR implementations introduce significant challenges, including synchronization, maintaining real-time determinism across the HW/SW boundary, and protecting shared memory.
A clear gap exists in the literature, as publicly documented and verified Co-TMR frameworks are rare.
The control and data migration mechanisms required for such asymmetric replication lack the formal verification support needed for safety-critical and redundant certification, marking a critical direction for future toolchain development.

\paragraph{Hierarchical TMR (H-TMR)} \label{par:hierarchical}
Also known as Multi-Layer TMR, introduces multi-level fault containment by recursively applying triplication across distinct architectural layers—e.g.
registers, subsystems, and interconnects. This strategy creates nested fault-containment domains and is common in aerospace and nuclear-grade systems where fault tolerance is demanded across data paths, control units, and bus fabrics.
The layers can span from physical resources to logical functions.
For instance, Yeh’s seminal “Triple-Triple” architecture~\cite{yeh_triple-triple_1996} for the Boeing 777 exemplifies this by layering redundancy across physical power supplies, communication paths (ARINC 629 buses), and dissimilar computational channels (using different microprocessors and compilers).
At a more granular level, Alagoz~\cite{alagoz_hierarchical_2008} presents a formal analysis, showing that adding hierarchical layers of voting—for example, by placing intermediate voters between replicated network nodes—can improve system reliability by orders of magnitude compared to a flat TMR structure.
A theoretical analysis projects that for a module with a 0.1\,\%\ error probability, a second-order TMR network reduces the system error probability to one in 37 billion.
Modern adaptive systems, such as the CPU-FPGA platform described by Mostl \emph{et al.}~\cite{mostl_self-adaptation_2019}, introduce another dimension of layering by separating a high-level control layer (on the CPU) from a reconfigurable hardware layer (in the FPGA), allowing redundancy to be managed dynamically across the software/hardware boundary.
This concept is formalized by Ding \emph{et al.}~\cite{ding_classification_2017}, who classify H-TMR as a combined design pattern where fault masking is applied progressively through a hierarchy.
While this layering makes the system highly robust by containing faults at the lowest possible layer, the main drawback is a significant increase in complexity.
Static, multi-layer designs are often based on worst-case assumptions, leading to poor resource utilization in nominal conditions~\cite{mostl_self-adaptation_2019}.
Furthermore, the design space development, verification and certification processes scale non-linearly due to the complex interactions between layers.
Despite these challenges, in ultra-critical domains like avionics, H-TMR remains a foundational strategy for fault-tolerant architecture.

\paragraph{Neural Network TMR}  \label{par:nn}
Integrates learning-based models into redundancy mechanisms, marking a recent innovation in fault-tolerant system design.
In contrast to classical \gls{TMR} methods that rely on rigid hardware duplication, AI-assisted redundancy leverages the predictive and diagnostic power of neural networks to enhance fault resilience and adaptability.

\paragraph{AI-Assisted TMR} \label{par:ai}
Where Ahmad \emph{et al.}~\cite{ahmad_modified_2025} propose a \gls{MTMR} architecture for \gls{TPMC}s that combines classical redundancy with a \gls{FDI} unit driven by an \gls{ANN}. The system uses the \gls{ANN} to diagnose open-circuit faults in real-time by monitoring current waveforms, enabling the system to isolate faulty switches and activate one of two redundant backups. This approach achieves a simulated system-level reliability of 99\% under both single and dual faults, significantly exceeding traditional \gls{DHR} or general \gls{DMR} designs.

This paradigm has also been extended to spaceborne systems. Shao \emph{et al.}~\cite{shao_research_2024} demonstrate that a \gls{CNN}-based classifier on a radiation-tolerant accelerator can achieve real-time fault masking using \gls{TMR}-isolated convolution blocks. Heavy-ion beam testing confirmed that classifier accuracy remained above 94\% even under soft-error injection, showing the potential of AI-assisted redundancy in harsh environments.

To balance efficiency and precision, other approaches focus on the selective protection of the neural networks themselves. Bertoa \emph{et al.}~\cite{bertoa_fault-tolerant_2023} developed an automated tool that identifies the most sensitive channels in a neural network and applies \gls{TMR} only to those critical computations, reducing hardware overhead while maintaining a user-defined level of accuracy. This strategy of tolerating imprecision in less significant \gls{ANN} layers directly mirrors the approximate computing frameworks (such as \gls{X-Rel}) previously discussed in Section~\ref{par:atmr}.

While promising, the field of \gls{AI}-assisted redundancy is still emerging. Current methods are often tailored to specific applications, such as power converters~\cite{ahmad_modified_2025}, or embedded within fixed accelerator designs~\cite{shao_research_2024}. A clear gap exists in the literature regarding unified frameworks that combine \gls{AI}-based diagnosis with dynamic, hardware-in-the-loop reconfiguration. Furthermore, the lack of established toolchains to synthesize \gls{ANN}-integrated TMR workflows for commercial \gls{FPGA}s or \gls{SoC}s limits scalability, representing a significant area for future research.

\paragraph{Radiation Hardened by Design (RHBD)} \label{par:rhbd}
While not a redundancy technique itself, \gls{RHBD} complements \gls{TMR} by structurally hardening logic against radiation-induced faults.
Instead of only relying on replicated logic, \gls{RHBD} uses transistor-level design techniques to prevent or suppress \gls{SEU} and \gls{SET} events directly on the silicon~\cite{schrape_design_2020}.
This reduces the fault rate of the underlying components, thereby increasing the overall reliability of a \gls{TMR} system.
Recent advances combine \gls{RHBD} with standard cell libraries, such as hardened \gls{TSPC} flip-flops that include internal \gls{SET} filters.
Schrape \emph{et al.}~\cite{schrape_design_2020} present a \gls{TMR} flip-flop using this approach, which saves area and improves performance compared to traditional D-latch-based designs.
This hybrid use of \gls{RHBD} and selective \gls{TMR} provides a robust solution for radiation-prone environments.
However, a gap remains in the literature regarding standardized methodologies for formally verifying the fault coverage of these combined RHBD-TMR systems, especially across different process nodes and radiation environments.
Furthermore, a cost-benefit calculation due to specialized hardware development and/or research has to be considered for \gls{RHBD}, \gls{TMR} or a combination.
\paragraph{Further TMRs} \label{par:ftmr}
Although this survey consolidates a wide range of \gls{TMR} techniques—from classical spatial and temporal schemes to mixed and AI-augmented variants—it is important to acknowledge that additional, niche or domain-specific \gls{TMR} architectures may exist beyond the categories addressed here.
In many cases, redundant mechanisms are developed under alternative names, embedded within proprietary systems, or applied in narrowly scoped environments, making systematic classification difficult.
Some redundancy strategies employ triplication only implicitly, others may blur the line between \gls{TMR} and N-Version programming, or be hidden under frameworks.
Moreover, due to the limited availability of public documentation in sectors such as defense, automotive, and spacecraft subsystems, it is likely that further practical \gls{TMR} schemes remain undocumented or inaccessible to the academic community.
The taxonomy proposed in this work is structured to provide clarity and coverage for well-documented and reproducible designs.

\subsection{Voters} \label{subsec:voters}

Voter logic is a critical component in any modular redundancy scheme, responsible for arbitrating between replicated modules to produce a single, valid output.
This is commonly achieved through majority voting or more advanced self-diagnostic consensus mechanisms~\cite{mitra_word-voter_2000, afzaal_self-checking_2018}.
While redundancy techniques themselves are well-documented, the classification of voter architectures in cooperation with redundancy scheme are not well documented.
To provide a clear framework for analysis and design, this survey categorizes voter techniques into five practical classes based on their primary operational goal and complexity:

\begin{itemize}
    \item \textbf{Basic Voters}: Fundamental mechanisms such as majority logic or word-level comparison, including bit-by-bit and word-wise implementations~\cite{mitra_word-voter_2000, balasubramanian_fault_2016, johnson_voter_2010}.
    
    \item \textbf{Optimization-Oriented Voters}: Designs that target resource efficiency by minimizing logic gate count (area), power, or timing overhead. This includes techniques like approximate voting (reducing precision to save area) or asynchronous synchronization, which is used for delay-insensitive coordination in temporal or clockless \gls{TMR} scenarios~\cite{vafaei_x-rel_2023, balasubramanian_area_2019}.
    
    \item \textbf{Spatial/Placement Voters}: Architectures that emphasize the physical topology, distribution, and isolation of the voting logic. These designs prioritize the separation of redundant domains to mitigate shared fault vulnerabilities like Multi-Bit Upsets (MBUs), rather than minimizing logic size~\cite{latif-shabgahi_integrating_1999, bergt_localized_2010}.
    
    \item \textbf{Adaptive/Dynamic Voters}: Reconfigurable voters that adapt their internal logic at runtime based on fault detection triggers or environmental feedback~\cite{arifeen_fault_2019, barbirotta_evaluation_2022}.
    
    \item \textbf{Feedback-Based Voters}: Systems integrating self-checking or error-correcting logic within the voter to detect internal voter failures, often utilizing feedback loops for validation~\cite{afzaal_self-checking_2018, reviriego_dmr_2016}.
\end{itemize}

\begin{figure}[htbp]
    \centering
\includegraphics[width=0.55\linewidth]{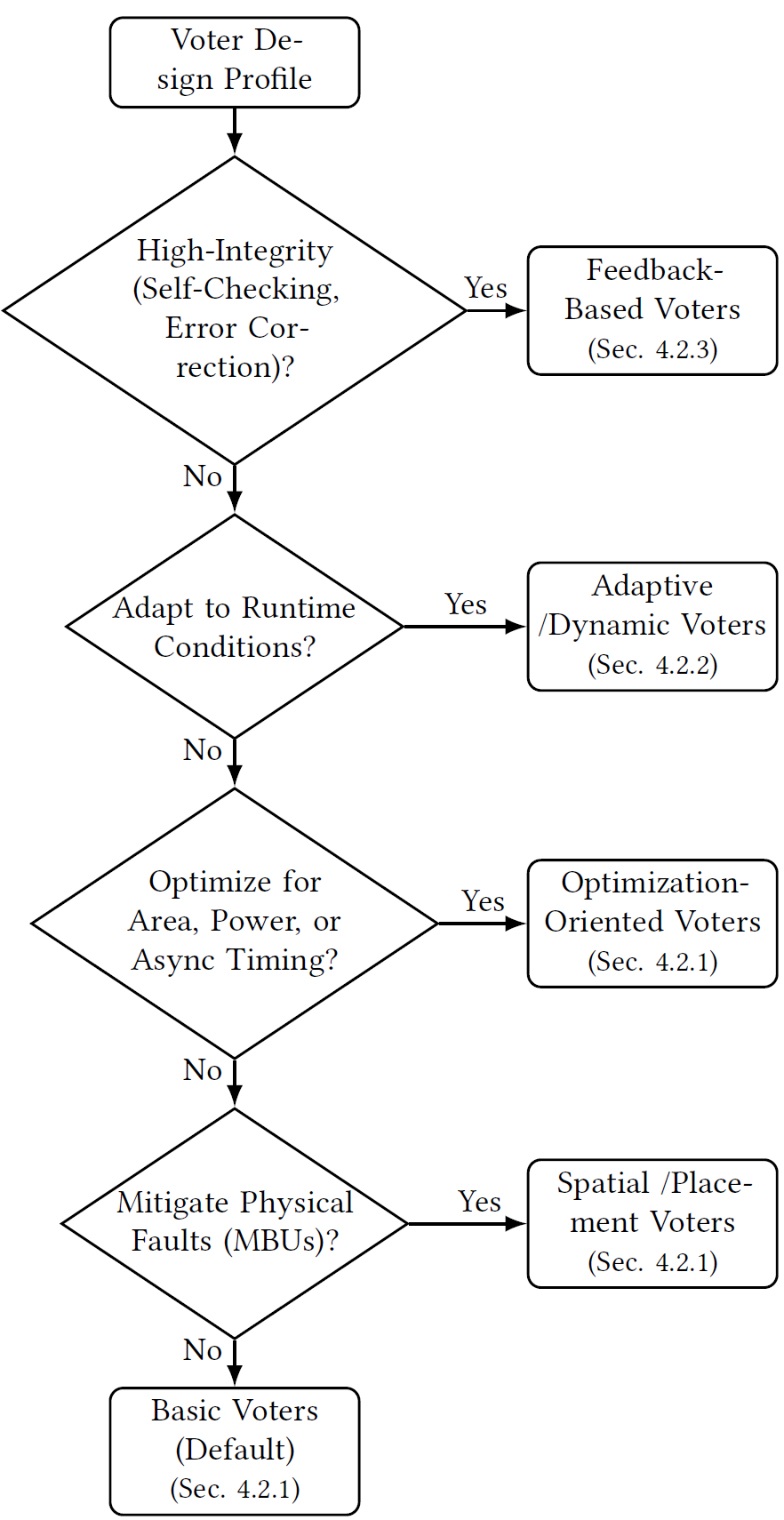}
\caption{Decision tree for selecting voter strategies.}
\Description{A flowchart starting with 'Voter Design Profile', progressing downwards through four decision diamonds. 'High-Integrity (Self-Checking, Error Correction)?' leads right to 'Feedback-Based Voters' on Yes, down on No. 'Adapt to Runtime Conditions?' leads right to 'Adaptive /Dynamic Voters' on Yes, down on No. 'Optimize for Area, Power, or Async Timing?' leads right to 'Optimization-Oriented Voters' on Yes, down on No. 'Mitigate Physical Faults (MBUs)?' leads right to 'Spatial /Placement Voters' on Yes, down to a final block 'Basic Voters (Default)' on No.}
\label{fig:voter_flow}
\end{figure}

\paragraph{Decision Guidance (Fig.~\ref{fig:voter_flow})}
Provides a logic route for identifying suitable voter architectures based on key system requirements and constraints.
A structured comparison is provided in the following subsections. At the time of writing, no single foundational paper for voter classification is available in current literature.
However, common structures are inferred from works such as Mitra and McCluskey~\cite{mitra_word-voter_2000} and Afzaal and Lee~\cite{afzaal_self-checking_2018}, which are frequently cited in the context of voter design and reliability.

\subsubsection{Basic Voters} \label{subsec:basic_voters}
These foundational designs implement the core majority logic required for fault masking without additional optimization or adaptive features.
They serve as the standard baseline for TMR systems, ranging from simple bit-wise gates to complete word-level comparators that ensure data coherence across the output vector~\cite{lyons_use_1962, mitra_word-voter_2000, johnson_voter_2010}.

\paragraph{Voter-less Redundancy}
Seeking to achieve fault tolerance without a centralized majority voter---which itself can act as a vulnerable single point of failure---this approach diverges from classical \gls{TMR}. As the necessity of an explicit 2-out-of-3 decision gate is removed, these methods generally fall under broader spatial redundancy rather than strict triplication.
Namazi \emph{et al.} propose one such strategy for defect-tolerant nano-architectures, aiming to mitigate the high defect rates in emerging technologies. Instead of relying on a discrete voter circuit to decide between redundant inputs, their approach utilizes massive redundancy, such as redundant interconnections, and leverages the inherent physical properties of the computing substrate. In such architectures, fault masking is achieved implicitly at the physical level (e.g., through signal superimposition or analog thresholding at wire intersections), where correct signals inherently overpower faulty ones. This implicit physical consensus completely avoids the logic overhead and reliability bottleneck of a complex, active voter circuit \cite{namazi_voterless_2008}.

\paragraph{Majority Logic (Decoding TMR)}
Often used as a synonym for the voter in a TMR system, is the fundamental mechanism that enables fault masking.
This logic decodes the redundant inputs to produce a single, correct output, assuming a fault affects no more than one module.
The classic implementation is a simple majority gate, which executes a "2-out-of-3" decision strategy (typically realized via the boolean function $Out = AB \lor BC \lor CA$) to filter out a single discordant signal~\cite{lyons_use_1962}.
However, as Balasubramanian and Prasad note, this classical voter can itself be a single point of failure.
Their work proposes an improved fault-tolerant majority voter that is more robust to faults occurring both internally and externally to the voter, addressing the limitations of simple majority decoding in nano-scale technologies~\cite{balasubramanian_fault_2016}.

\paragraph{Bit-by-bit Voter}
Is the most conventional implementation of majority logic, where an independent voter gate is applied to each bit of the output data bus.
As Mitra and McCluskey describe, this is the "conventional TMR system" design \cite{mitra_word-voter_2000}.
While simple and fast, this approach can be vulnerable to faults that cause compensating errors across different bits, potentially leading to a corrupt but voted correct output \cite{siewiorek_reliability_1974}.
Ló \emph{et al.} note that this scheme is the most basic, simple and quick but may fail to ensure data integrity if multiple bits are affected in a way that escapes simple majority logic at each bit position \cite{lo_towards_2014}.

\paragraph{Word-Level Voter} \label{subsubsec:word_level_voter}
Proposed by Mitra and McCluskey, compares entire output vectors from replicated modules instead of voting on a bit-by-bit basis.
This design enhances data integrity by ensuring that a majority of modules agree on the complete output word in the same clock cycle.
If no two output words match, the voter flags an error, preventing the propagation of an incorrect state that might arise from compensating faults across different bits.
This approach is particularly suited for high-throughput, synchronous systems with wide data buses, where the temporal coherence of the entire data vector is critical \cite{mitra_word-voter_2000}.

\paragraph{Redundant Voters}
Are used to mitigate the voter itself as a single point of failure.
This technique, as implemented by Afzaal and Lee, involves replicating the voting circuit itself, often alongside a self-checking or consensus mechanism to arbitrate between the voter copies.
Their analysis demonstrates that using redundant voting copies can significantly increase the overall system reliability compared to an unprotected simplex voter \cite{afzaal_self-checking_2018}.
Recent work by Deepa M. and Augusta Sophy Beulet P. also explores improvised voter architectures with redundancy to reduce area overhead while improving fault-masking capabilities \cite{m_improvised_2022}.

\paragraph{Output Majority Voters}
Are placed at the final outputs of a redundant system, as opposed to intermediate stages.
This is the classic TMR configuration described by Lyons and Vanderkulk \cite{lyons_use_1962}.
While this approach masks faults that occur anywhere within the triplicated modules, it can be vulnerable to fault accumulation in sequential systems.
Johnson and Wirthlin note that in designs with feedback paths, voters are required in all such paths to ensure proper synchronization, not just at the primary outputs \cite{johnson_voter_2010}.

\paragraph{Optimization-Oriented Voters} \label{subsec:Optimization-Oriented_Voters}
These architectures prioritize resource efficiency and timing determinism, modifying standard voting logic to reduce area and power consumption or to safely bridge asynchronous clock domains.
The primary goal is to minimize the implementation overhead of the voter while maintaining valid consensus, often through approximate computing or delay-insensitive protocols~\cite{vafaei_x-rel_2023, balasubramanian_area_2019}.

\paragraph{Clock Domain Crossing Voters}
Are used, when redundant modules operate in different or asynchronous clock domains, a standard majority voter is insufficient.
The system is vulnerable to "asynchronous sampling uncertainty," where signal skew between the triplicated paths causes replicas to arrive in the receiving clock domain on different cycles~\cite{li_synchronization_2010}.
This creates a synchronization hazard where a simple \gls{SEU} can defeat the \gls{TMR} mask~\cite{li_synchronization_2010, johnson_voter_2010}.
A \gls{CDC} voter mitigates this by incorporating robust synchronizer circuits (e.g. multi-stage flip-flops) into its inputs, ensuring data is stable in the voter's clock domain before comparison.
While essential, this adds latency and introduces the synchronizers themselves as potential points of failure~\cite{li_synchronization_2010}.

\paragraph{Synchronization Voters} \label{par:sync_voters}
Are a general class of voters designed to manage and ensure timing correctness in redundant systems.
The concept was first formalized as "synchronization voting," which uses mutual feedback to correct for drift without a common clock~\cite{davies_synchronization_1978}, with specific designs depending heavily on the signaling convention (e.g. pulse, level, or transition)~\cite{mcconnel_synchronization_1981}.
In modern synchronous systems, the term "synchronization voter" often refers to voters placed in feedback paths to prevent "persistent errors" by re-synchronizing the state of the \gls{TMR} domains each cycle~\cite{johnson_voter_2010}.
In true asynchronous (clockless) systems, synchronization is achieved with \gls{QDI} voters, which use handshaking protocols to coordinate the arrival of data.
As demonstrated by Balasubramanian \emph{et al.}, these \gls{QDI} voters are inherently tolerant to timing variations and can be optimized for area, making them suitable for robust, low-power asynchronous designs~\cite{balasubramanian_area_2019}.

\paragraph{Reducing Voters}
Aim to minimize the significant area, power, and delay overhead associated with voter logic, especially in wide-data-path designs.
This is a key goal of approximate reliability frameworks like X-Rel (see Section~\ref{par:atmr}), proposed by Vafaei \emph{et al.}
By relaxing the precision of the voter to only compare a subset of the most significant bits, the size and energy consumption of the voter logic can be drastically reduced.
This optimization is highly effective in error-tolerant applications where the full precision of the output is not critical \cite{vafaei_x-rel_2023}.

\paragraph{Partitioning Voters}
Or adaptable-width voters, offer a compromise between bit-by-bit and word-level voting.
As described by Ló \emph{et al.}, this technique groups bits into sets or partitions, with a separate voter function for each.
This approach can improve data integrity over bit-voting by ensuring coherence within a partition, while being less complex and resource-intensive than a full word-voter.
This design space allows for optimization based on system constraints, such as the number of modules and the acceptable probability of corrupt outputs in the presence of multiple errors \cite{lo_towards_2014}.

\paragraph{Spatial/Placement Voters} \label{subsubsec:spatial_placement_voters}
Addressing physical vulnerabilities such as Multi-Bit Upsets (MBUs), these strategies focus on the topology and isolation of the voter itself.
Techniques include distributing decision logic across the design or integrating it with module inputs to prevent localized single points of failure from disabling the voting mechanism~\cite{latif-shabgahi_integrating_1999, bergt_localized_2010}.

\paragraph{Integrated Voters} \label{par:int_voters}
Or hybrid voters, combine traditional fault masking with external self-diagnosis mechanisms. Unlike feedback-based voters that monitor their own internal health, integrated voters utilize diagnosis information provided by the redundant modules themselves. Latif-Shabgahi \emph{et al.} propose such a voter that integrates this module-level "health status" into the voting algorithm. By considering both the output "value" and the self-reported health of each module, the voter can make a more informed decision, particularly in cases where a simple majority vote would fail or be ambiguous. This approach captures advantages from both fault masking and fault detection/isolation, enabling the voter to select more correct results than logic based on output values alone \cite{latif-shabgahi_integrating_1999}.

\paragraph{Distributed Voter TMR}
Voting logic is not centralized but is instead distributed throughout the design, often placed at the inputs of replicated modules or at intermediate pipeline stages.
This contrasts with \gls{LTMR}, where replicas are placed closely together \cite{bergt_localized_2010}.
Distributing the voters helps to prevent fault propagation between redundant domains and can improve synchronization in systems with feedback loops \cite{johnson_voter_2010}.
Aysan \emph{et al.} extend this concept with their "Voting on Time and Value" (VTV) strategy, which distributes voters that check both the correctness of the value and its arrival time, making it suitable for real-time distributed systems \cite{aysan_vtv_2008}.

\subsubsection{Adaptive/Dynamic Voters} \label{subsec:Adaptive_Dynamic_Voters}
Designed for systems requiring runtime flexibility, these mechanisms alter their decision logic or precision in response to changing environmental conditions or detected fault patterns.
They are essential for mixed redundancy schemes where the system must shift between high-reliability and high-performance modes~\cite{arifeen_fault_2019, barbirotta_evaluation_2022}.

\paragraph{Adaptive/Dynamic Voters}
Are mechanisms that can alter their internal logic or behavior in response to runtime conditions.
Advancements in \gls{ATMR} voter design, for example, have produced fault-tolerant voters that improve reliability by incorporating both detection and recovery stages~\cite{arifeen_fault_2019}.
\gls{ATMR} approaches, including the prominent \gls{FAC} framework~\cite{balasubramanian_fac_2023}, bridge the gap between design efficiency and fault tolerance, offering a viable path for certification-constrained, low-power applications in embedded domains.

\paragraph{Hybrid Voters}
Combine different voting strategies or integrate voting with other fault-tolerance techniques.
For instance, a voter in a dynamic \gls{TMR} system, such as that proposed by Barbirotta \emph{et al.}, must adapt its logic.
It may operate in a simple comparison \gls{DTMR} mode for fault detection and only engage its full three-input majority logic when a fault is flagged and a recovery thread is activated~\cite{barbirotta_evaluation_2022}.
This concept also includes voters that integrate diagnosis information (e.g. a "health status") from the modules themselves to make a more informed decision, capturing advantages from both fault masking and fault detection~\cite{latif-shabgahi_integrating_1999}.

\paragraph{Selective TMR Voters}
System, where voters are not applied universally but are placed only at the outputs of selectively triplicated critical modules.
This strategy is based on pre-deployment sensitivity analysis to identify the most vulnerable components~\cite{samudrala_selective_2004}. As demonstrated by Bertoa \emph{et al.}
in their work on fault-tolerant neural network accelerators, \gls{STMR} hardens only these high-impact channels.
The voters are therefore also selective, placed strategically to mask faults from critical paths while non-critical data passes through non-redundant logic, balancing reliability and resource overhead~\cite{bertoa_fault-tolerant_2023}.

\paragraph{Register Voters with Refresh} \label{par:register_refresh_voters}
Is a technique, which implements a synchronous, temporal refresh mechanism directly at the register level.
Reviriego \emph{et al.} propose the \gls{DMR}+ design where the values of replicated flip-flops are periodically overwritten by their majority state (or a known-good copy)~\cite{reviriego_dmr_2016}.
This active refresh strategy uses voters to correct transient \gls{SEU}s in sequential elements with significantly less area overhead than a full spatial \gls{TMR} implementation.
A related concept is the "synchronization voter" placed in all feedback paths, which passively ensures that the state of \gls{TMR} domains remains synchronized and prevents persistent errors from accumulating in registers~\cite{johnson_voter_2010}.

\subsubsection{Feedback-Based Voters} \label{subsec:feedback-based_Voter}
To ensure high integrity, these designs augment the majority vote with internal self-checking mechanisms or feedback loops. Crucially, these mechanisms are focused on the voter itself: they allow the system to detect or correct failures internally within the voter logic, preventing the decision unit from becoming a silent single point of failure~\cite{afzaal_self-checking_2018, reviriego_dmr_2016}.

\paragraph{Voter with Feedback (Inside or Outside)}
Augment traditional majority voters by embedding error-detection mechanisms.
These systems can detect voter malfunctions, caused by transient faults or internal logic corruption, and either flag an error or switch to a redundant internal copy.
Afzaal and Lee present a self-checking voter for \gls{FPGA}s that uses redundant internal voter copies and an arbiter to switch between them upon fault detection, improving reliability by approximately 26\,\%\ over a simplex voter~\cite{afzaal_self-checking_2018}.
Other hybrid voters use feedback from the modules themselves, such as a "health status", to make a more informed decision than logic based on output values alone~\cite{latif-shabgahi_integrating_1999}.

\paragraph{Self-Checking / Self-Diagnostic Voters}
Are designed to be "Totally Self-Checking (TSC)" or "Strongly Fault-Secure (SFS", as theoretically defined by Nicolaidis and Courtois~\cite{nicolaidis_self-checking_1986}. In addition to performing the majority vote, these voters continuously check their own internal logic for faults. As implemented by Afzaal and Lee, such a voter uses redundant internal copies and a comparator. If an internal fault is detected, the voter signals an error, preventing it from producing an incorrect output that "appears" correct. This is crucial for high-integrity systems, as it makes the voter itself fault-secure~\cite{afzaal_self-checking_2018}.

\paragraph{Error Detection / Correction}
Many advanced voters incorporate \gls{EDC} logic directly, turning them into active repair mechanisms. For the sake of completeness, it is essential to acknowledge that foundational \gls{EDC} heavily relies on established information-theoretic standards from the literature, such as Hamming codes \cite{hentschke_analyzing_2002}, Reed-Solomon codes \cite{gaoNewAlgorithmDecoding2003}, or Cyclic Redundancy Checks \cite{castagnoliOptimizationCyclicRedundancycheck1993} (\gls{CRC}s). While these classical methods are traditionally used to protect data integrity within memory arrays or communication buses, advanced voter architectures extend these principles directly into the redundancy arbitration process. The voter presented by Arifeen \emph{et al.} for \gls{ATMR} systems, for example, includes both detection and recovery stages to manage faults within the logic paths~\cite{arifeen_fault_2019}. Similarly, the \gls{DMR}+ technique from Reviriego \emph{et al.} uses a voter that first detects a mismatch between two modules and then actively corrects the faulty register by overwriting it with the value from the correct copy~\cite{reviriego_dmr_2016}.

\section{Conclusion and Future Work}
This survey has provided a comprehensive classification of redundancy techniques and voter architectures, analyzing the current state-of-the-art and clarifying fragmented terminology. This concluding chapter summarizes the key findings, synthesizes practical design implications, and proposes concrete directions for future research. These insights aim to guide practitioners in their design choices and researchers toward impactful new areas of investigation.

\subsection{Summary and Comparative Insights}
The field of fault-tolerant electronic systems is vast and mature, yet it is simultaneously hindered by significant terminological fragmentation that obscures the relationships between functionally similar techniques. A primary finding of this survey is that the academic and industrial landscape is replete with techniques that are functionally identical but described with different nomenclature. For example, methods classified under Temporal Redundancy such as \gls{RMT} \cite{chen_task_2016}, \gls{TMT} \cite{barbirotta_evaluation_2022}, and Software-level \gls{TMR} \cite{janson_software-level_2018} all leverage the re-execution of software tasks to mask transient faults, yet they are rarely cross-referenced.

To address this, this work establishes a unified taxonomy by organizing redundancy methods into three primary categories—Spatial, Temporal, and Mixed—and introducing a novel five-class taxonomy for voter logic. This framework aims to demystify the complex landscape of fault tolerance, providing a foundational reference that enables engineers to navigate the trade-offs between reliability, cost, and performance.

\subsection{Practical Design Implications}
Based on the decision flowcharts and comparative analysis presented in this survey, several practical guidelines emerge for system architects:

\begin{itemize}
    \item \textbf{Safety-Critical Baseline:} For high-stakes applications (e.g., avionics, nuclear control) where certification is paramount, full Spatial \gls{TMR} (Section \ref{sec:spatial_redundncy}) remains the non-negotiable approach. The resource cost is justified by the requirement for immediate, in-line fault masking.
    
    \item \textbf{Resource-Constrained Optimization:} For cost-sensitive \gls{MCU}s or small \gls{FPGA}s, Temporal Redundancy (Section \ref{sec:temporal_redundancy}) offers a viable path to high reliability. This is effective provided the application's real-time constraints can tolerate the latency penalty of re-execution.
    
    \item \textbf{Dynamic Adaptability:} For complex \gls{SoC}s operating in variable radiation environments, the state-of-the-art is moving toward Mixed or Adaptive Redundancy (Section \ref{sec:mixed_redundnacy}). Techniques like \gls{ATMR} are particularly promising for error-tolerant applications (e.g., \gls{AI} acceleration) where graceful degradation is acceptable \cite{vafaei_x-rel_2023, arifeen_approximate_2020}.
    
    \item \textbf{Voter Selection:} The voter must not be an afterthought. While a simple Bit-by-bit Voter is fast, it is highly vulnerable to \gls{MBU}s. A Word-Level Voter provides significantly more robust data integrity \cite{mitra_word-voter_2000}. For high-integrity systems, a Feedback-Based Voter is essential to prevent the voter itself from becoming a single point of failure \cite{afzaal_self-checking_2018}.
\end{itemize}

\subsection{Open Challenges}
Despite decades of research, significant challenges remain. 

\paragraph*{System-Level Gaps and Proprietary Barriers}
A primary challenge is the "black box" nature of industrial implementations. While System-Level \gls{TMR} is known to be used in systems like the Boeing 777 \cite{yeh_triple-triple_1996} and in \gls{ISO} 26262 compliant automotive designs \cite{noauthor_iso_nodate}, the specific details of inter-board synchronization, voter design, and fault isolation are often proprietary. This lack of public-domain data makes it difficult for the academic community to analyze and improve upon real-world, high-integrity systems.

\paragraph*{Implementation Complexity}
The engineering effort required for Mixed Redundancy, particularly Reconfigurable \gls{TMR} (Section \ref{sec:mixed_redundnacy}), remains exceptionally high. As highlighted by Vipin and Fahmy \cite{vipin_fpga_2018}, the absence of unified, high-level \gls{API}s in commercial \gls{FPGA} toolchains for managing dynamic partial reconfiguration remains a major bottleneck, forcing designers to rely on low-level, error-prone bitstream manipulation.

\paragraph*{Emerging Physical Threats}
While vulnerabilities have already been well-documented at the 28\,nm node, the continuous shrinking of technology nodes into the deep sub-nanometer regime drastically amplifies the threat of Multi-Bit Upsets (\gls{MBU}s). In these highly dense modern architectures, a single particle strike is increasingly likely to corrupt multiple adjacent logic cells simultaneously, defeating any \gls{TMR} implementation where redundant domains are not sufficiently isolated \cite{quinn_domain_2007}. This renders simple Localized \gls{TMR} insufficient for high-radiation environments, demanding strictly spatially distributed implementations which, however, are not always feasible in area-constrained devices.

\subsection{Future Research Directions}
Based on the gaps identified, four primary directions are proposed for future research:

\begin{itemize}
    \item \textbf{Standardization of Terminology:} A concerted effort is needed to standardize the terminology of fault tolerance. This survey proposes the classification herein as a starting point to facilitate clearer communication and more direct comparison between methods.
    
    \item \textbf{\gls{MBU} Mitigation Models:} Research must focus on quantifiable models that link advanced process technologies (e.g., scaling from 28\,nm down to modern deep sub-nanometer nodes) to the required physical separation distances. This is crucial for developing layout-aware \gls{TMR} strategies that are verifiably resilient to \gls{MBU}s in highly dense, next-generation devices \cite{sterpone_layout-aware_2011}.
    \item \textbf{Cross-Domain Learning:} Methodologies are needed to incorporate findings from high-radiation testbeds. Analyzing radiation-hardened implementations used in high-energy physics (e.g., CERN) can inform broader design strategies for other radiation-intensive environments, such as aerospace or medical applications \cite{coronetti_radiation_2023, jung_strahlenqualifizierung_2023}.
    
    \item \textbf{AI and High-Level Toolchains:} The role of \gls{AI} should move beyond simple diagnostics toward predictive fault-tolerance \cite{shao_research_2024}. Concurrently, the development of high-level toolchains is critical to abstract the complexity of partial reconfiguration, making dynamic fault tolerance accessible to a wider range of designers.
\end{itemize}

\subsection{Final Thoughts}
Redundancy is not a "one-size-fits-all" solution. It is a complex and dynamic field involving a series of critical trade-offs between reliability, performance, power, and cost. As information processing systems in all domains—from automotive to aerospace—move toward greater autonomy, their dependence on fail-operational computing will only increase. The ability to intelligently select, implement, adapt, and verify these redundant techniques will be paramount. This survey has aimed to build, as well as enhance, the foundational map for the field and future developments of redundancy techniques.

\begin{acks}
The authors would like to gratefully acknowledge the ongoing collaboration and fruitful discussions with our partners at the European Organization for Nuclear Research (CERN), the Bergische University of Wuppertal, and the ATLAS Inner Tracker (ITk) collaboration. Their invaluable insights into the practical challenges of deploying microelectronics in high-radiation environments, as well as their expertise in particle detector data acquisition systems, have significantly enriched the practical context of this survey. Furthermore, we extend our gratitude for the broader institutional support and research facilities that made this comprehensive study possible.
\end{acks}

\bibliographystyle{ACM-Reference-Format}
\bibliography{references}

\end{document}